\documentclass[tightenlines,twocolumn,prl,notitlepage,nofootinbib,preprintnumbers,longbibliography]{revtex4-1}
\usepackage{graphicx}
\usepackage{hyperref}
\hypersetup{colorlinks=true, linkcolor=blue, urlcolor=blue, citecolor=blue, final}
\usepackage{tabularx}
\usepackage{multirow}
\usepackage{array}
\usepackage[percent]{overpic}
\usepackage{float}
\usepackage[inline]{asymptote}
\usepackage{rotating}
\usepackage{epsfig}
\usepackage{xcolor}
\usepackage{amsmath}
\usepackage{amssymb}
\usepackage{bm}
\usepackage{pict2e}

\usepackage{tikz}
\usetikzlibrary{positioning}
\usetikzlibrary{matrix}
\usepackage[normalem]{ulem}
\usepackage{verbatim}

\begin{document}

\title{Enhancing Angular Sensitivity of Segmented Antineutrino Detectors for Reactor Monitoring}

\author{Brian~C.~Crow}
\author{Max~A.~A.~Dornfest}
\author{John~G.~Learned}
\author{Jackson~D.~Seligman}
\author{Nathan~S.~Sibert}
\author{Jeffrey~G.~Yepez}
\affiliation{Department of Physics and Astronomy, University of Hawai'i at M\={a}noa}
\author{Viacheslav~A.~Li}
\affiliation{Lawrence Livermore National Laboratory}

\date{\today}

\begin{abstract}
We present a potential improvement over the standard method developed to determine antineutrino directionality in inverse-beta-decay detectors. The previously developed method for quantifying directionality in monolithic and segmented detectors may be ambiguous in methodology. In this paper, we present a directionality algorithm and include error analysis. We have developed an algorithm based on a measure of ``distance'' between two matrices. We report findings for our research in reactor-antineutrino directionality, and emphasize that the algorithm has broad applications whenever one desires computationally efficient 2D pattern-matching.
We treat data from Lithium-6-doped detector segments in the form of a matrix.
The validation of our algorithm boils down to comparing a Monte Carlo generated ``empirical'' data set to a simulated data set. 
The empirical data set is generated for a particular orientation of the neutrino beam. 
We identify an optimal segmentation scale in the low-count regime.
We also discuss the shortcomings of the conventional method and how this knowledge can be applied to segmented detectors, hybrid designs, and generalized validation, agnostic to the physics of detector design.
\end{abstract}
\maketitle 

\section{Introduction to antineutrino directionality in segmented detectors for inverse beta decay}
Antineutrinos from beta decay provide a unique probe of nuclear processes in environments that are otherwise difficult to access.  
These isotopes which undergo beta decay are found in nuclear reactors, nuclear explosions~\cite{Bernstein:2009ab, Foxe:2020vtb, 10.1063/5.0263319}, spent nuclear fuel~\cite{PhysRevApplied.8.054050,Radermacher:2022pnt, doublechoozcollaboration2025measurementneutrinoemissionsspent}, and the earth's mantle and crust~\cite{Watanabe:2014, Leyton:2017tza}. 
Since they only interact with the weak force and gravity, one of the advantages of using neutrinos for nuclear reactor monitoring is that it is nearly impossible to develop an effective shielding for these particles. 
Most neutrino detection applications focus on the detection of antineutrinos from fission for this reason. 
A use for developing such a detector and refining its angular sensitivity is that we can locate the antineutrino source, with a higher level of precision, reducing the area encompassed by the angular distribution, thereby reducing the uncertainty of the location of the source. The practicality of this application can be improved by improving the directional sensitivity of the inverse beta decay (IBD) detection channel. Several approaches to improving this sensitivity through the selection of appropriate geometry have been explored~\cite{Duvall:2024cae}, but there remains room for further improvement by refining the analysis technique of reconstructing antineutrino direction from IBD kinematics. 

In the IBD interaction,
\[
^1\mathrm{H}^+ + \bar\nu_e \to n + \;e^+,
\]
a proton in a hydrogen atom transforms into a neutron, while the antineutrino is converted into a positron via W-boson exchange. The threshold energy for this reaction is approximately 1.8~MeV, arising from the proton–neutron mass difference of about 1.3~MeV and the positron mass of 0.511~MeV. Most of the antineutrino energy is carried away by the emitted positron.
The direction of the outgoing neutron is correlated with the direction of the incoming antineutrino. This forward-backward asymmetry for the neutron was originally observed in the Gosgen experiment~\cite{Zacek:1984qi} (and later in the Palo Verde experiment~\cite{PaloVerde_direction}), and the neutron directionality has been later observed in the Bugey experiment~\cite{Cussonneau:1992lua}, theoretically analyzed by Beacom and Vogel~\cite{PhysRevD.60.053003}, and further experimentally investigated in the Chooz~\cite{PhysRevD.61.012001} and DoubleChooz experiments~\cite{Caden:2012bm}, including on hydrogen~\cite{Roncin:2014adw}. Recent results from the PROSPECT experiment~\cite{PhysRevD.111.032014}, along with theoretical investigations conducted by our group~\cite{Duvall:2024cae}, further contribute to this field.
The directional nature of the IBD reaction prompts a discussion on the potential utility of detectors that utilize this process. Before proceeding with detailed simulations, we will discuss the underlying physics and perform basic analytical calculations to gain insight into the order of magnitude for timing and distance related to neutron captures. Some of the formalism can be found in the original Reines and Cowan article~\cite{Reines_RevSciIns1954}, as well as in one of our earlier publications~\cite{mTC:2016yys}.

\subsection{Analytical Evaluation of IBD}\label{subsec:analytical_eval}
A positron of a few MeV energy will deposit most of its energy in the form of ionizing radiation before annihilation with an electron in the material. Annihilation in-flight is possible, but is usually of negligible effect --- on the order of a few percent~\cite{2017EPJWC.14608010S}. After reaching thermal energy and before annihilation, there is also an approximately 50\% chance of forming a short-lived positron-electron system (positronium)~\cite{DoubleChooz:2014iyf}. 

Assuming a stopping power of 2~MeV~cm$^{-2}$~g$^{-1}$, the range of a 1-MeV positron in plastic with density 1.1~g/cm$^3$ can be estimated as
\[
\Delta x \simeq \frac{\Delta E}{\rho S} \simeq 5~\text{mm}.
\]
Thus, a typical IBD positron will travel on the order of a centimeter before annihilation. In organic scintillators, the light yield is approximately $10^4$ photons per MeV of electron-equivalent energy. The scintillation light from the positron is directly detectable using appropriate photomultipliers. The two 511-keV annihilation gamma rays travel farther and deposit their energy away from the primary IBD vertex.
 
Depending on the segment size, the positron may highlight multiple segments; however, for simplicity of the calculation in this paper, we assume the prompt (i.e. positron) event as a single-segment event. This will become most apparent in our results section where we present plots of both total and usable events. For the finely segmented SANDD-like detector, an algorithm may be used in the future to better estimate the location of the IBD vertex ~\cite{Sutanto:2021xpo}.

After being produced in the IBD process, neutrons lose energy primarily through elastic scattering on the nuclei of the scintillator. The energy lost per collision depends on the target nucleus mass; light nuclei such as hydrogen are especially effective in slowing down neutrons due to their mass being comparable to that of the neutron. Hydrogen-rich scintillators are therefore efficient moderators. For scattering on hydrogen, a neutron loses on average about half of its energy per collision. The number of collisions needed to slow a neutron from initial energy $E_i$ to final energy $E_f$ is approximately
\[
n = \log_2\left(\frac{E_i}{E_f}\right).
\]
For a 4-keV neutron (typical for IBD) to thermalize to 0.025~eV, about 17 collisions are required.

Thermal-neutron captures on lithium-6,
\[
^6\mathrm{Li}(n,\alpha)^3\mathrm{H},
\]
are highly localized, since the charged reaction products have short ranges of order $10~\mu\mathrm{m}$ and no gamma rays are emitted (unlike captures on hydrogen, boron-10, or gadolinium). The reaction releases 4.78~MeV, shared between the alpha and the triton according to their masses:
\[
E_\alpha = \frac{m_T}{m_\alpha + m_T} E_{\text{tot}} = \frac{3}{7} E_{\text{tot}} \approx 2.08~\text{MeV},
\]
with the remaining 2.73~MeV carried by the triton. Due to quenching, a 2.08-MeV alpha particle produces light equivalent to roughly 0.5~MeV of electron-equivalent energy. In a segmented scintillator, the capture energy is fully contained within a single segment, which is advantageous for reconstructing the capture vertex, since the spatial resolution is set by the segment size and position.

\begin{figure*}[ht!]
\centering
\includegraphics[width=0.98\linewidth]{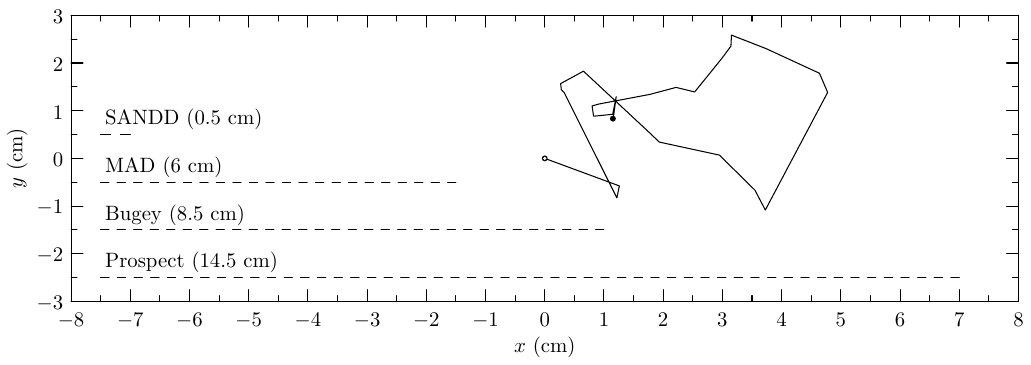}
    \caption{To-scale diagram showing a characteristic neutron $xy$-trajectory projection from IBD vertex to capture location, as well as dimensions of segments in four 2D-segmented detectors: SANDD, MAD, Bugey, and PROSPECT. If printed on a letter/A4 format (21.6-mm-wide paper), the lines would correspond to proper size of those detector segments. This particular event if started from the center of either MAD or Prospect segment would have the capture event in the same segment as the prompt.}
    \label{fig_segmentation_scaling}
\end{figure*}

The capture time constant can be estimated as $(\sigma n v)^{-1}$, where the cross section $\sigma = 941 \pm 3$ barns  (or $(9.41 \pm 0.03) \times 10^{-22}$ cm$^2$) \cite{osti_6280345}, $n$ is the number density, and $v$ is the thermal-neutron velocity (2200 m/s for $mv^2 /2 = kT$ at room temperature).
Based on the equation, the capture time constant is approximately 45.5 and 9.2 microseconds, respectively for 0.1\% and 0.5\% loading. Add about 10 microseconds for the neutron to reach thermal energy. Using realistic number densities and macroscopic cross section $\Sigma = \sigma n$, capture fraction on lithium-6 (941 barns) relative to the captures on hydrogen (0.33 barns) can be estimated as $\Sigma(^6\mathrm{Li}) / \Sigma (^1\mathrm{H})$. For 0.1\% (0.5\%) doping levels, for each capture on hydrogen there are 6.75 (34.5) captures on $^6$Li. When neutron captures on hydrogen, it turns into deuterium and emits 2.2-MeV gamma radiation. Although it could also be used for directionality (Double Chooz paper further explored it), in this paper we focus on $^6$Li due to the localized capture location.
The mean free path is about 1~cm (for elastic scatter off hydrogen $1/ (n_H \sigma_{es})$, $\sigma_{es}$ is about 20 barns). The average distance traveled by the neutron before thermalization is about 4 centimeters ($\sqrt{17} \; \times$ 1 cm).
This means for sufficiently large segment sizes, $\mathcal{O}$(10~cm), the 
majority of capture events would be within the same segment as the prompt event (simulated event shown in Fig.~\ref{fig_segmentation_scaling}).

\subsection{IBD detectors with ${}^6$Li as a dopant}
Starting with the Bugey 3 experiment~\cite{Bugey3_1995, Bugey3_1996}, antineutrino detectors and prototypes that use segmented scintillators doped with lithium-6 include SANDD, MAD, ROADSTR~\cite{6t97-9bpb}, PROSPECT, and the originally proposed NuLat experiment~\cite{lane2015newtypeneutrinodetector}.
SOLID and CHANDLER also use ${}^6$Li, but only on the outer surfaces of their segments (3D segmentation). This confinement of the ${}^6$Li dopant to the surfaces of the segments in these detectors makes the neutron detection layer much thinner. This thin neutron detection layer contradicts the implicit assumption of isotropy in the binning scheme introduced in Figs. \ref{fig_segmentation_scaling} and \ref{fig:phi_def}. Modifying the binning and the directionality algorithm to address these anisotropies is not within the scope of this work.
The detectors and their main characteristics are listed in Table~\ref{tab_Li6_detectors}. The ongoing development of ${}^6$Li-doped 3D-printable plastics further motivates this directionality study~\cite{KIM2023168537, Barr:2025fxg}.
Fig.~\ref{fig:Bugey_Prospect_data} highlights the neutrino directionality in segmented $^6$Li detectors using data from Bugey 3 and PROSPECT experiments.

\begin{figure}[!h]
\begin{tikzpicture}[scale=1.25]

\draw[thick] (0,0) grid (3,3);

\node at (0.5,2.5) {---};
\node at (1.5,2.5) {7.3\%};
\node at (2.5,2.5) {---};

\node at (0.5,1.5) {11.3\%};
\node at (1.5,1.5) {49.8\%};
\node at (2.5,1.5) {11.9\%};

\node at (0.5,0.5) {---};
\node at (1.5,0.5) {19.7\%};
\node at (2.5,0.5) {---};

\draw[thick,->,black,>=stealth] ({1.5+0.25*cos(110)}, {-0.5+0.25*sin(110)}) -- ({1.5+0.25*cos(-70)}, {-0.5+0.25*sin(-70)});

\node at (1.5, -1) {Bugey: $\theta_\nu \simeq -70^\circ$};

\begin{scope}[shift={(3.5,0)}]
\draw[thick] (0,0) grid (3,3);

\node at (0.5,2.5) {---};
\node [align=center] at (1.5,2.5) {3800\\7.5\%};
\node at (2.5,2.5) {---};

\node [align=center] at (0.5,1.5) {1718\\3.4\%};
\node [align=center] at (1.5,1.5) {39276\\77.7\%};
\node [align=center] at (2.5,1.5) {3766\\7.5\%};

\node at (0.5,0.5) {---};
\node [align=center] at (1.5,0.5) {2009\\4.0\%};
\node at (2.5,0.5) {---};

\draw[thick,->,black,>=stealth] ({1.5+0.25*cos(221)}, {-0.5+0.25*sin(221)}) -- ({1.5+0.25*cos(41)}, {-0.5+0.25*sin(41)});

\node at (1.5, -1) {Prospect: $\theta_\nu = 41^\circ$};
\end{scope}

\end{tikzpicture}
\caption{Bugey 3 and PROSPECT directional data. For Bugey, only percentages were originally reported \cite{Cussonneau:1992lua}, and only the inner-segment data is shown here. Report baselines and active reactor core sized. For Bugey, the detector was almost directly below the power reactor ~\cite{PhysRevD.111.032014, Jayakumar:2024job}. The neutrino directions are shown by the corresponding arrows.}
\label{fig:Bugey_Prospect_data}
\end{figure}

\begin{figure}[ht]
    \includegraphics[width=1\linewidth]{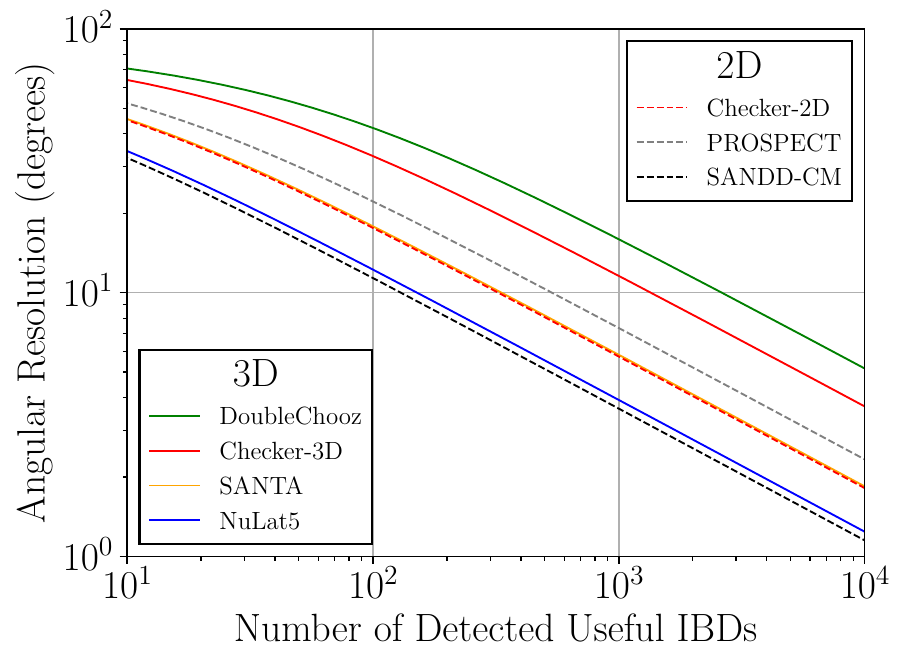}
    \caption{Angular uncertainty calculated using the conventional ``Chooz" approach. A selected number of segmented detectors along with Chooz is highlighted. Useful IBD events are those that have prompt and delayed signals in separate segments. This result is taken from our previous study~\cite{Duvall:2024cae}.}
    \label{fig_old_method}
\end{figure}

To be more sensitive to directionality, it is beneficial that neutrons capture in a different segment with respect to the IBD vertex. 
Higher doping level leads to shorter path length between reaching the thermal energy and the capture.
Therefore, it is desirable to have segment scale on the order of the prompt-delayed distance, as we will see in the angular uncertainty results. The fraction of useful events for directionality, as reported in our previous paper can be estimated as follows:
SANDD --- $\sim$100\%, NuLat --- 86\%, PROSPECT --- 39\%.
The angular uncertainty is shown in Fig.~\ref{fig_old_method} using the conventional method.

\begin{table*}
    \centering
 \begin{tabular}{|l|c|c|c|c|r|r|c|}\hline
    Experiment & Seg. size, cm$^3$ & Number of seg. & L/P & \% w. & Active mass, kg & $ N_{\mathrm{H}} $ & Ref. \\\hline
    Bugey 3 & 8.5 $\times$ 8.5 $\times$ 85 & 98 (7 $\times$ 14) & L & 0.15 & 518 & $ 3.74 \times 10^{28} $ & \cite{Bugey3_1996}\\
    NuLat$^{*}$ & 5 $\times$ 5 $\times$ 5 & 3375 (15 $\times$ 15 $\times$ 15) & P & 0.1 & 435 & $ 2.21 \times 10^{28} $ & \cite{lane2015newtypeneutrinodetector}\\
    Prospect & 14.5 $\times$ 14.5 $\times$ 117.6 & 154 (11 $\times$ 14)  & L & 0.082 & 3655 & $ 2.06 \times 10^{29} $ & \cite{Ashenfelter_2019} \\
    miniChandler & 6.2 $\times$ 6.2 $\times$ 6.2 &  320 (8 $\times$ 8 $\times$ 5) & P & --- & 79 & $ 3.99 \times 10^{27} $ & \cite{Haghighat:2018mve} \\
    Solid & 5 $\times$ 5 $\times$ 5  & 12800 (50 $\times$ 16 $\times$ 16) & P & --- & 1651 & $ 8.38 \times 10^{28} $ & \cite{Abreu_2021}\\
    SANDD$^{*}$ & .5 $\times$ .5 $\times$ 40 & 64 (8 $\times$ 8) & P & 0.1 & 0.7 & $ 3.35 \times 10^{25} $ & \cite{Sutanto:2021xpo} \\
    ROADSTR & 5.5 $\times$ 5.5 $\times$ 50 & 36 (6 $\times$ 6) & P & 0.1 & 56 & $ 2.85 \times 10^{27} $ & \cite{6t97-9bpb} \\
    MAD-2D & 6 $\times$ 6 $\times$ 100 & 64 (8 $\times$ 8) & P & 0.1 & 238 & $ 1.21 \times 10^{28} $ & \cite{bowden_mobile_2023} \\
    MAD-3D & 5 $\times$ 5 $\times$ 2.5  & 6400 (16 $\times$ 16 $\times$ 25)  & P & --- & 413 & $ 2.09 \times 10^{28} $ & \cite{bowden_mobile_2023} \\
    \hline
\end{tabular}
    \caption{Experiments with $^6$Li as a neutron-capture agent. Asterisk indicates proposed or prototype. For SANDD, the smallest segment size is listed. Solid, miniChandler, MAD-3D use sheets of LiZnS-based scintillator, the IBD ``target'' scintillator itself is not doped, so all the neutron captures are on the surfaces of the segments. ``L/P'' stands for Liquid or Plastic form. Active mass and $ N_{\mathrm{H}} $ are computed assuming PVT for plastic ($\rho$ = 1.032 g/cm$^{3}$, $ w_{\mathrm{H}} = 10/118 $), EJ-309-like DIPN for PROSPECT liquid ($\rho$ = 0.96 g/cm$^{3}$, $ w_{\mathrm{H}} = 20/212 $), LAB-like liquid for Bugey 3 ($\rho$ = 0.86 g/cm$^{3}$, $ w_{\mathrm{H}} = 0.121 $).} 
    \label{tab_Li6_detectors}
\end{table*}

Conventional IBD direction reconstruction in scintillator relies on the prompt-to-delayed displacement vector and typically estimates the angular uncertainty via straightforward uncertainty propagation. In a segmented detector, this framing can create the impression that the achievable directional precision is primarily dictated by the segment pitch (or, in a monolithic detector, by the vertex resolution), and that arbitrarily fine segmentation could in principle drive the uncertainty to very small values. That interpretation is incomplete because it treats the displacement as though it were limited mainly by instrumental resolution, rather than by the underlying event physics and transport.

In reality, the IBD neutron is emitted with a substantial intrinsic kinematic spread relative to the incoming antineutrino direction (of order of tens of degrees) even before any scattering, diffusion, or thermalization. Subsequent moderation and capture further de-correlate the capture point from the initial neutron direction, so the residual directionality is only weakly preserved in the final capture location. Standard estimators based on simple displacement averaging, or on component-wise Gaussian assumptions and error propagation, do not explicitly incorporate this intrinsic broadening and can therefore yield misleadingly optimistic uncertainty estimates, especially at low event counts. Motivated by this, we develop an alternative approach that abandons closed-form, resolution-dominated error propagation in favor of a purely data-driven (template or pattern-matching) inference of the incident direction using the full displacement pattern. 
A practical advantage is that the method remains applicable in the low-statistics regime where Gaussian approximations and conventional uncertainty estimates break down, which is relevant for geo-neutrino and core-collapse supernova pointing, as well as in scenarios with multiple or spatially separated neutrino sources (for example, multiple reactor cores or distributed geo-neutrino emission).

\subsection{Motivation for a new algorithm to estimate neutrino directionality}

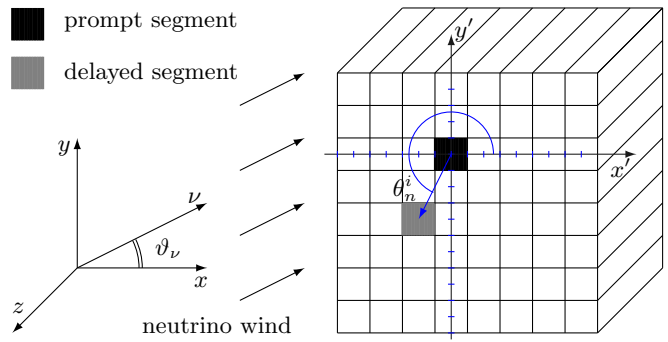
\begin{figure}[ht]
        \setlength{\unitlength}{.1\linewidth}
\begin{picture}(10,5)
        \multiput(5,0)(.5,0){9}{\color{black}\line(0,1){4}} 
        \multiput(5,0)(0,0.5){9}{\color{black}\line(1,0){4}} 

        \multiput(5,4)(.5,0){9}{\color{black}\line(1,1){1}} 
        \multiput(9,0)(0,.5){9}{\color{black}\line(1,1){1}} 

        \put(10,1){\line(0,1){4}}
        \put(6,5){\line(1,0){4}}

          \thicklines
        \multiput(6.5, 2.50)(.025,0){20}{\color{black}\line(0,1){0.5}} 
        \multiput(0, 5)(.025,0){20}{\color{black}\line(0,-1){0.5}} 
        
        \multiput(6, 1.50)(.025,0){20}{\color{gray}\line(0,1){0.5}} 
        \multiput(0, 4.25)(.025,0){20}{\color{gray}\line(0,-1){0.5}} 
          \thinlines

        \put(0.8,4.7){prompt segment}
        \put(0.8,3.9){delayed segment}

        \put(1,1){\vector(-1,-1){1}} 
        \put(1,1){\vector(0,1){2}} 
        \put(1,1){\vector(1,0){2}} 

        \put(1,1){\vector(2,1){2}} 
        \put(1,1){\arc[0,26]{1}} 
        \put(1,1){\arc[0,26]{.95}}

        \put(0,0.3){$z$}
        \put(0.7,2.8){$y$}
        \put(2.8,.7){$x$}
        \put(2.7,2){$\nu$}
        
       \put(6.75,-0.1){\color{black!90}\vector(0,1){4.7}} 
        \multiput(6.7,0.)(0,.25){16}{\color{blue}\line(1,0){0.1}}
        \put(4.8,2.75){\color{black!90}\vector(1,0){4.8}} 
        \multiput(5,2.7)(.25,0){16}{\color{blue}\line(0,1){0.1}} 

        \put(6.8,4.55){\color{black}$y'$}
        \put(9.2,2.4){\color{black}$x'$}
        \put(5.85,2.1){\color{black}$\theta_n^i$}

        \put(6.75,2.75){\color{blue}\arc[0,245]{0.65}} 
        \put(6.75,2.75){\color{blue}\vector(-1,-2){0.5}}

        \multiput(3.5, 0.5)(0,1){4}{\color{black}\vector(2,1){1}} 
        
       \put(2,0){neutrino wind} 
        \put(2.2,1.2){$\vartheta_\nu$}
\end{picture}
        \caption{Diagram explaining the prompt-delayed segment geometry, as well as neutrino and axes orientation.
        In this work we assume that neutrino ``wind'' is in $xy$ plane, at an angle $\nu$ with the $x$ axis.
        The prompt and delayed segment in this diagram shown for illustrative purposes; the actual separation depends on the segment size --- for large segments, prompt and delayed events would happen in the same segment. It is also illustrated that capture is not necessarily aligned with the neutrino beam (only on average it is the case as discussed in this paper).
        To study the angular dependence we vary the angle of neutrino with respect to the detector's $x$ and $y$ axes (i.e. rotation of the detector along $z$ axis).}
        \label{fig:phi_def}
\end{figure}

Let us consider a 2D detector, schematically shown in Fig. \ref{fig:phi_def}. In the conventional formalism, the procedure for determining angular uncertainty $\Delta\varphi$ can be summarized as follows \cite{Duvall_2022}
\begin{equation}
      \Delta \varphi = \arctan \left( \frac{P/l}{\sqrt N} \right),
      \label{eq:deltaphi}
\end{equation}
where $\Delta\varphi$ is the $1\sigma$ angular uncertainty on reconstructed direction to antineutrino source;
$P$ is the mean position resolution, which can be estimated as the average location resolution;
$l$ is the mean distance between prompt and delayed events, 
$N$ is the number of IBD events used for direction reconstruction.

\begin{figure}[ht]
    \centering    
    \includegraphics[width=0.495\linewidth]{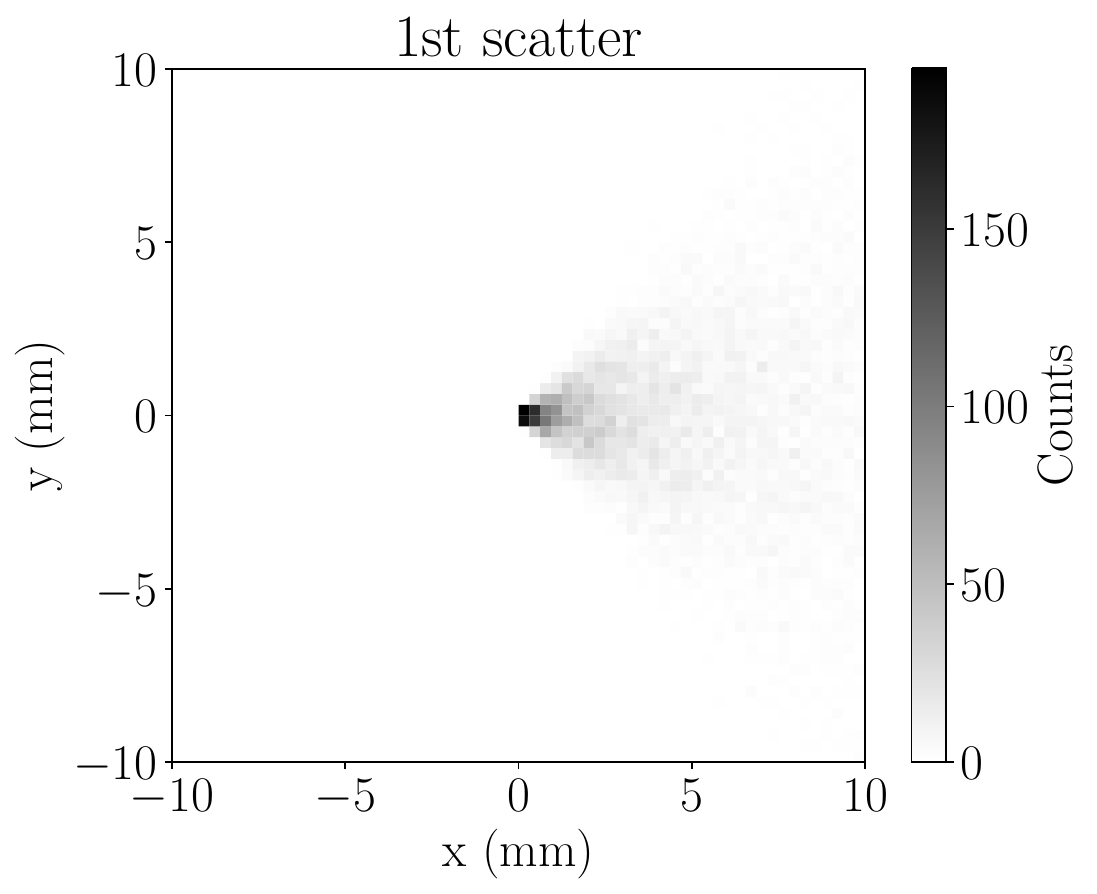}
    \includegraphics[width=0.485\linewidth]{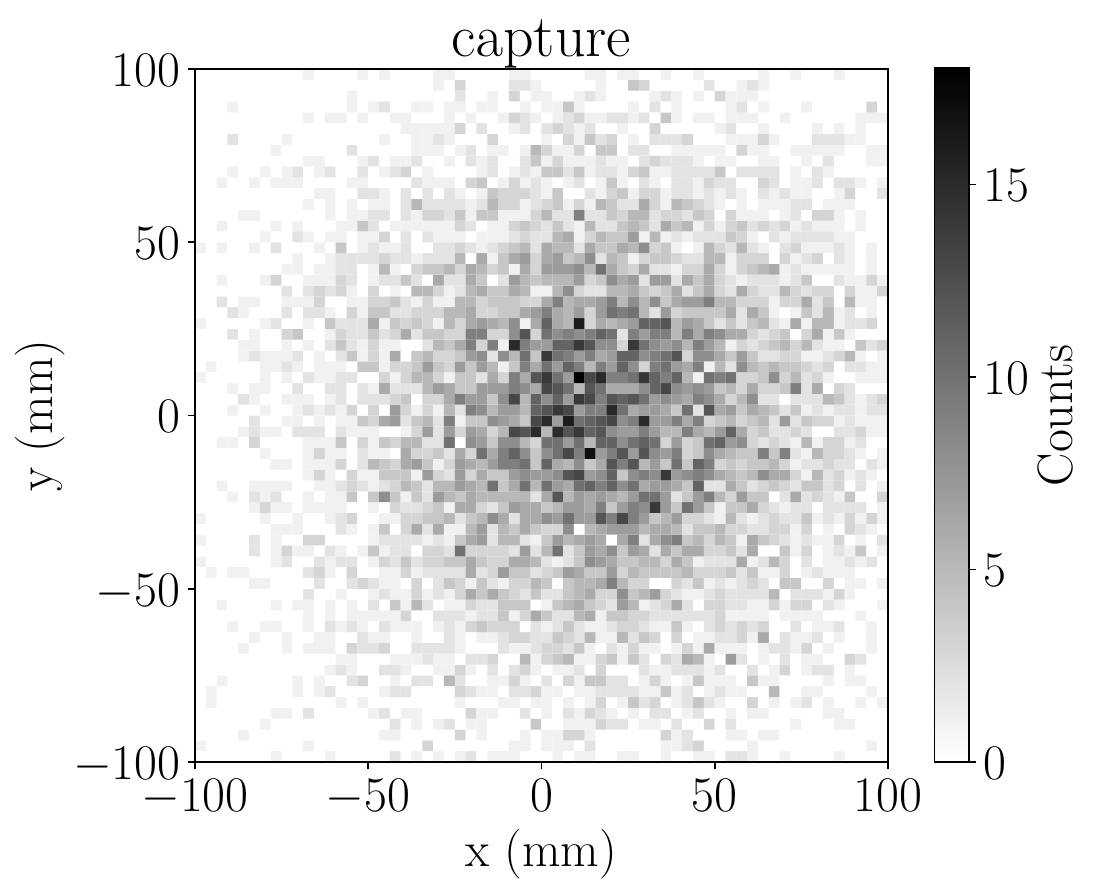}\\
    \includegraphics[width=0.49\linewidth]{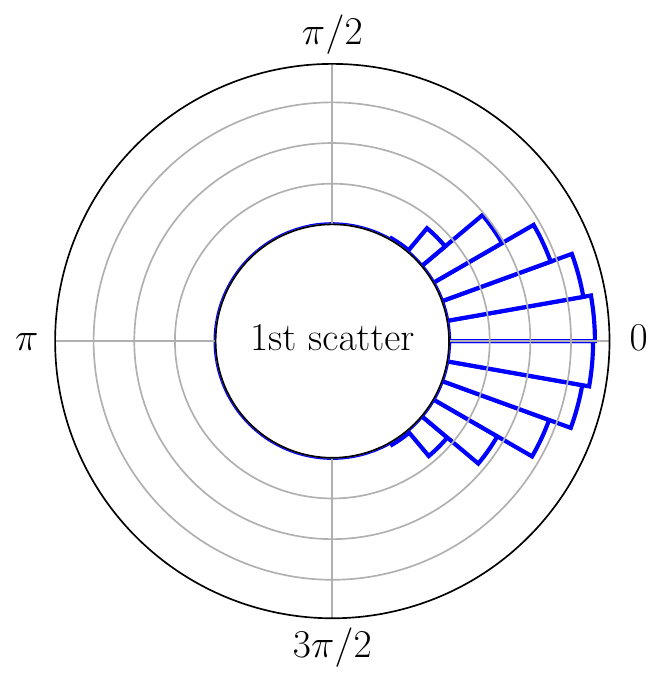}
     \includegraphics[width=0.49\linewidth]{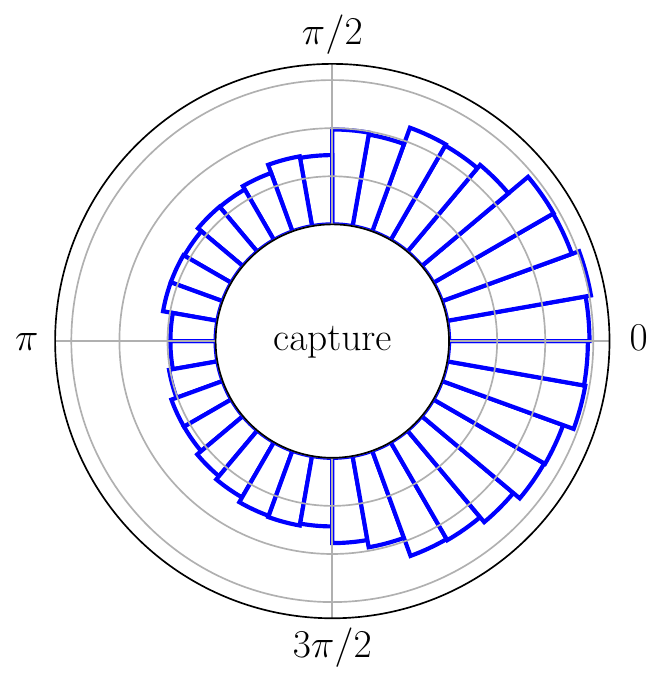}
    
    \caption{Top: Spatial distribution ($xy$ projection) for simulated IBD neutrons: first scatters (left) and  captures (right) locations, relative to the IBD vertex placed at the origin. Bottom: Polar histograms of neutron first-scatter (left) and capture  (right) locations from IBD simulated in RATPAC2. These distributions are for a 10k event run with 0.1\% ${}^6$Li loading. Note a clear bias toward the $+x$-direction  --- the direction of incoming neutrino is along positive $x$ axis (zero angle).}
    \label{fig:cloudplots}
\end{figure}

\begin{figure}[ht!]
    \centering
     \begin{overpic}[width=1.\linewidth]{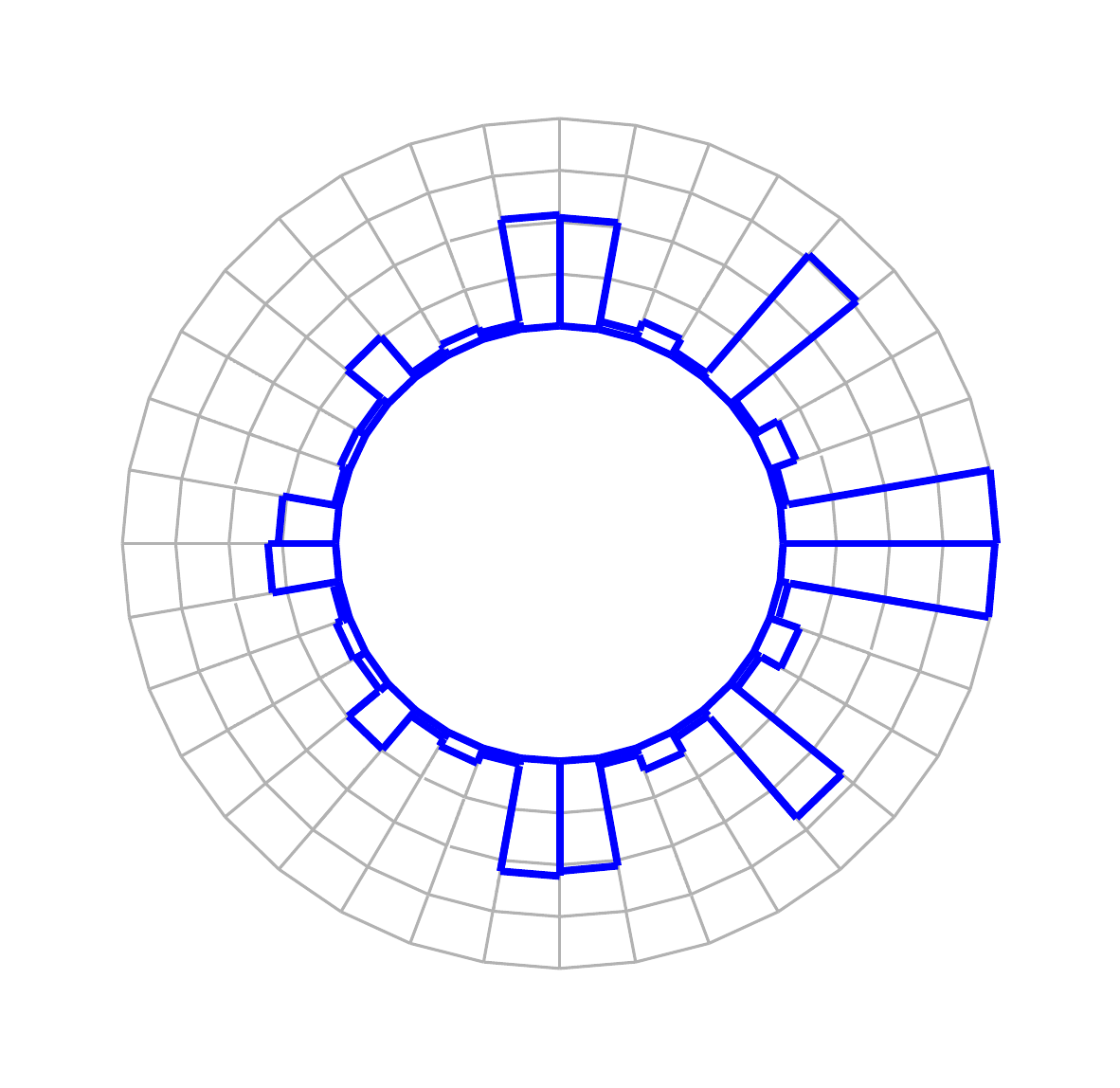}
        \put(91,47){\color{black}0}
        \put(46,90){\color{black}$\pi/2$}
        \put(5,47){\color{black}$\pi$}
        \put(45,5){\color{black}$3\pi/2$}
        \put(41,47){\color{black}Segmented}
    \end{overpic}
    \caption{Angular distribution of neutron captures in a segmented detector. Figure taken from our previous study~\cite{Duvall_2022}.}
    \label{fig_paperA_radar_segmented}
\end{figure}
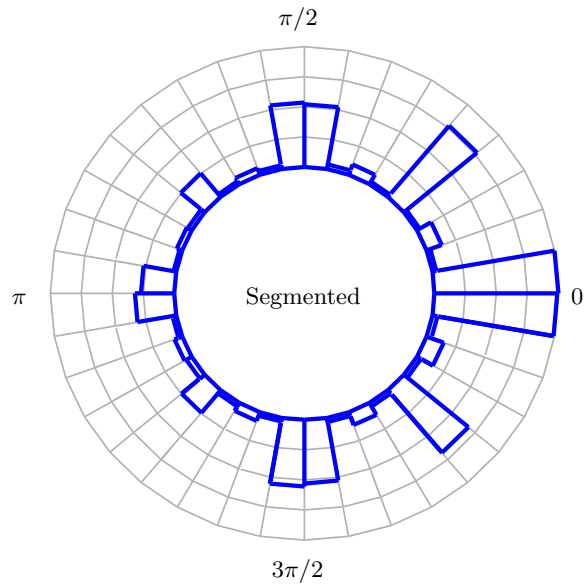

In this conventional formalism, if the mean position resolution $P$ goes to 0, the mathematics in Eq.~\ref{eq:deltaphi} inappropriately suggests that there is a corresponding ideal vertex reconstruction, with the lower limit on the angular uncertainty being an asymptote to zero. This is not realistic since neutron capture locations still have some finite non-zero volume (spatial spread). With accurate vertex reconstruction (i.e. low $P$) the formalism can return an arbitrarily small angular uncertainty --- even including the case for a small number of events. Additionally, there is a large spread ($\sim$ 26 degree) in the initial neutron direction. This is equally true for captures: they exhibit a typical drunken-walk with incredibly high variability (including back-scattering), as demonstrated in the spatial and angular distributions in Fig.~\ref{fig:cloudplots}. 
If we take the individual event direction to be from the center of the prompt  cell to the center of some nearby capture cell (adjacent, diagonal, or further) relative to the prompt cell, this would lead to a very inhomogeneous distribution.
The angular distribution for neutrons \textit{in segmented detectors} does not have a solved analytical form (for example, shown in Fig.~\ref{fig_paperA_radar_segmented}). As a result, the validity of using gaussian statistics directly over these angular distributions is questionable. 
However, with sufficiently large event numbers, experiments --- starting with Chooz and most recently with PROSPECT --- were able to reconstruct the direction of the reactor, down to the few-degree level.

The questions before us are: First, when does the old formalism begin to make sense? Second, what is the proper way to quantify the angular uncertainty $\delta \theta$ such that it correctly quantifies the error or uncertainty on that predicted angular uncertainty. 

In a study conducted in 2007 for DoubleChooz~\cite{DC2007}, exploring two-core scenario, when it came to distinguishing between the two cores it became difficult when the reactors are close together (separation angles below 45 degrees or so).
The two reactors become smudged together, and therefore the notion of resolution in Eq.~\ref{eq:deltaphi} is artificial. 
Another motivation is that we would like an algorithm  to be robust enough to handle large distributed sources of interest for geo-neutrino applications~\cite{Watanabe:2014}.
There have been new proposals for quasi-segmented detectors such as LiquidO~\cite{consortium2023probingearthsmissingpotassium} that would be sensitive to both reactor and geo-neutrino applications. 

In the old formalism, delayed events occurring in the same segment as the prompt events are not included in the analysis as they carry zero information. In the formalism we propose, these events should be included by default, but could be excluded if needed or assigned a different weight.
In general, two quantities can have the same value but different uncertainties depending on how they were measured. If the measurements are correlated, the way their uncertainties combine in calculations is affected, but the individual uncertainties themselves are not required to be equal or unequal solely due to correlation. In other words, the uncertainty on the angle of neutrino is not equal to the uncertainty on the angle of the neutron $\Delta \theta_\nu \ne \Delta \theta_n$. Therefore, angular resolution on the direction of the reactor is not  uncertainty on the neutron-capture displacement. However, in this method based on pattern-matching, it is possible to extract the angular resolution on the reactor direction since the simulations are performed by varying the neutrino direction (and essentially extracting it, not the neutron-capture displacement/direction).

\begin{table}[h]
    \centering
    \begin{tabular}{|c|c|c|c|c|c|c|c|}
        \hline
        \%\;wt\;$^6$Li & C & H & O & N & $^6$Li & $\Sigma_{\mathrm{Li}} / \Sigma_\mathrm{H}$ &  $\epsilon$ \\
        \hline
        0.1 & 4.70 & 4.48 & 0.207 & 0.0890 & 0.0106 & 6.75 & 87\% \\
        \hline
        0.5 & 4.13 & 4.37 & 0.632 & 0.0890 & 0.0528 & 34.5 & 97\% \\
        \hline
    \end{tabular}
\caption{Number densities for the scintillator modeled in this study: 0.1\% of $^6$Li loading by weight, the values are given in units of $10^{22}$ atoms per cm$^3$. $\Sigma_{\mathrm{Li}} / \Sigma_\mathrm{H}$ is the ratio of macroscopic cross sections for capture on $^6$Li and H; while $\epsilon$ is the fraction of capture on $^6$Li.}
    \label{tab:li6_loading}
\end{table}

\begin{figure}
\centering
\includegraphics[width=0.98\linewidth]{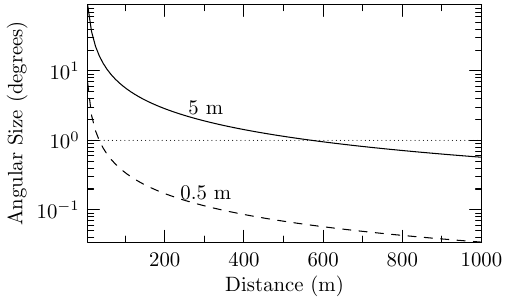}
    \caption{Angular size of a reactor active core as a function of distance. Two distinct cores shown for comparison (a 5-m radius and $<$0.5-m corresponding to a typical power reactor and research or SMR cores). Analytical calculation $\alpha = 2 \;\mathrm{atan}(R/d)$, where $R$ core radius, $d$ --- distance between reactor and detector. Detector is assumed point-like and the angular size is not power-weighted (in reality there will be more neutrinos at the centerline, than at the outliers). A sub-1-degree scale (dotted line) is approached at distances $\gtrsim 50$~m and $\gtrsim 500$~m, respectively.}
    \label{fig_reactor_angular_size_analytic}
\end{figure}

In this paper, we focus on the specifics of the algorithm and have made the following assumptions regarding our sources and signals. First, the \(1/r^2\) effect is neglected, meaning that the same neutrino flux is assumed for all segments. This assumption is not realistic for near-field detectors, as the \(1/r^2\) effect is observed~\cite{Jayakumar:2024job} \cite{AlekseevI.2025Ltrr}; specifically, a higher inverse beta decay event rate as exhibited by segments that are closer to the reactor.
Second, the angular sizes of the reactor and detector are assumed to be negligible. Consequently, it is assumed that all neutrinos have momenta in the same direction, resulting in no angular spread --- essentially treating the neutrinos as a collinear beam or ``wind.'' Figure~\ref{fig_reactor_angular_size_analytic} illustrates the angular size of typical reactor cores as a function of distance.
Third, events involving escaping neutrons are not included. This includes both neutrons that exit the detector and later return, as well as those that do not capture within the detector.

\renewcommand{\arraystretch}{1.0} 





\section{Simulating IBD positron and neutron pairs}
Reactor Antineutrino Tool Plus Additional Code (RAT-PAC version 2) is used to simulate IBD events~\cite{osti_2377202}.
RAT is a simulation and analysis package developed with GEANT4, ROOT, and C++, initially created for the Braidwood collaboration in early 2000s~\cite{BOLTON2005166}, a UK-US experiment to measure the mixing angle $\theta_{13}$, which later merged to some extent with the DayaBay collaboration~\cite{HAHN2006134}.
The code is publicly available via the GitHub repository (link in the reference cited above).
RAT-PAC is now utilized by various particle physics experiments, including SNO+ and Theia. 
The newer version (2) is primarily developed by LBL and LLNL.
It integrates simulation and analysis into a unified framework, enabling easy access to the same detector geometry and physics parameters in both detailed analysis and simulations. Adhering to the ``AMARA'' principle --- As Microphysical as Reasonably Achievable --- RAT tracks each photon through highly detailed detector geometries, employing standard GEANT4 or custom physics processes, while fully modeling PMTs and allowing the propagation of detected photons to simulations of front-end electronics and data acquisition systems.

\begin{figure}[ht]
    \begin{overpic}[width=1.0\linewidth]{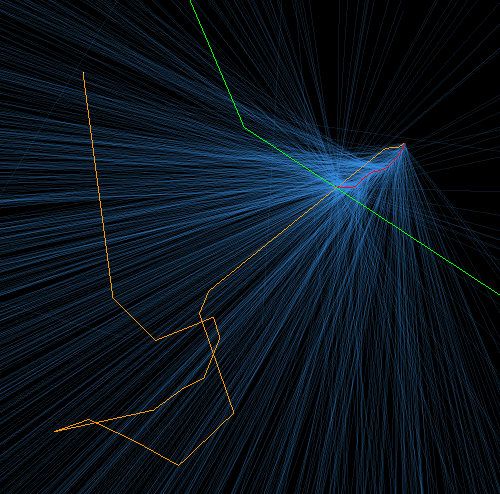}
    
    \put(16.5,90){\color{white}\vector(0,-1){5}}
    \put(0,93){\fcolorbox{white}{white!30}{\parbox{70pt}{neutron capture ${\color{orange}n} \; {^6}\mathrm{Li} \to {^3}\mathrm{H}\;\; {^4}\mathrm{He}$}}}
    
    \put(67,45){\color{white}\vector(0,1){15}}
    \put(57,41){\fcolorbox{white}{white!30}{\parbox{50pt}{positron\\ annihilation ${\color{red}e^+} e^- \to {\color{green}\gamma} \; {\color{green}\gamma}$}}}

    \put(81,86){\color{white}\vector(0,-1){15}}
    \put(65,85){\fcolorbox{white}{white!30}{\parbox{70pt}{inverse $\beta$ decay $\bar\nu_e \; ^1\mathrm{H} \to {\color{red}e^+ } \;{\color{orange}n}$}}}
    \end{overpic}
    \caption{Rendering of RAT-PAC 2 IBD visualization. Positron track in red, neutron track --- orange, gamma-rays --- green, Cherenkov photons --- cyan. Scintillation light turned off for clarity. Cherenkov photons indicate the direction of the positron, which traveled from the top right to the bottom left in this perspective (neutrino, not shown, is from top right to bottom left.}
    \label{fig_RATPAC_IBD_visual}
\end{figure}

\begin{figure}[ht!]
    \centering
    \begin{overpic}[width=1\linewidth]{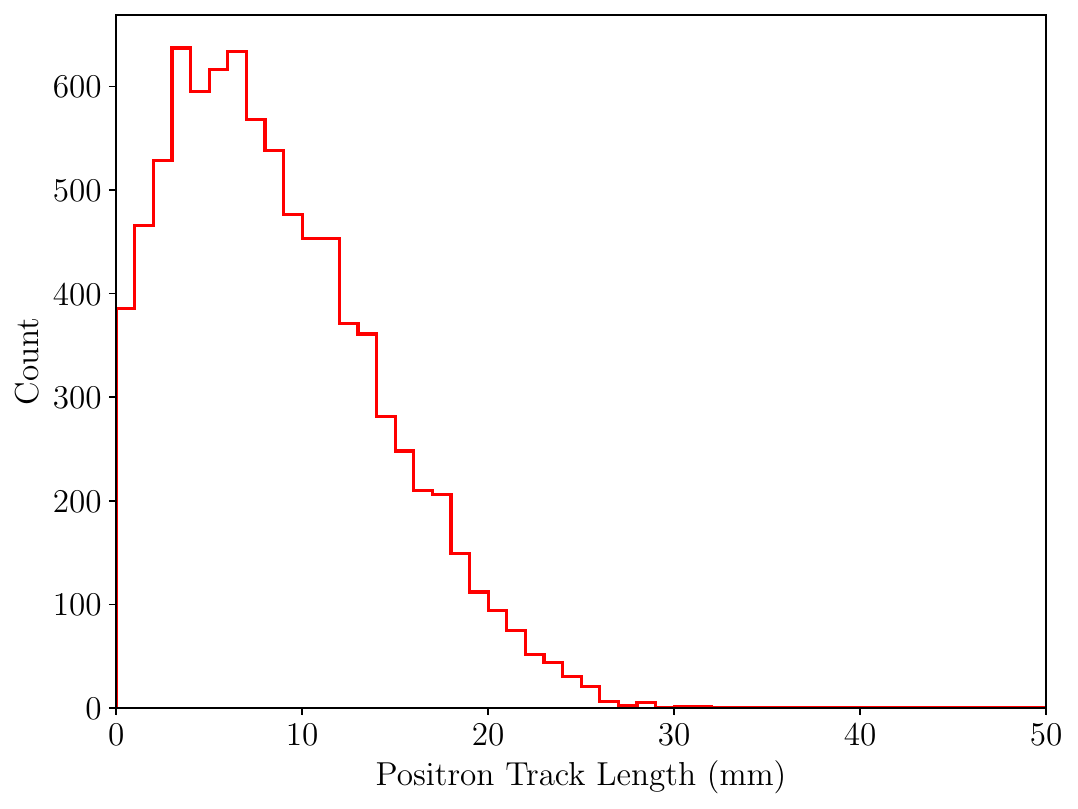}
        
    \end{overpic}
    \begin{overpic}[width=1\linewidth]{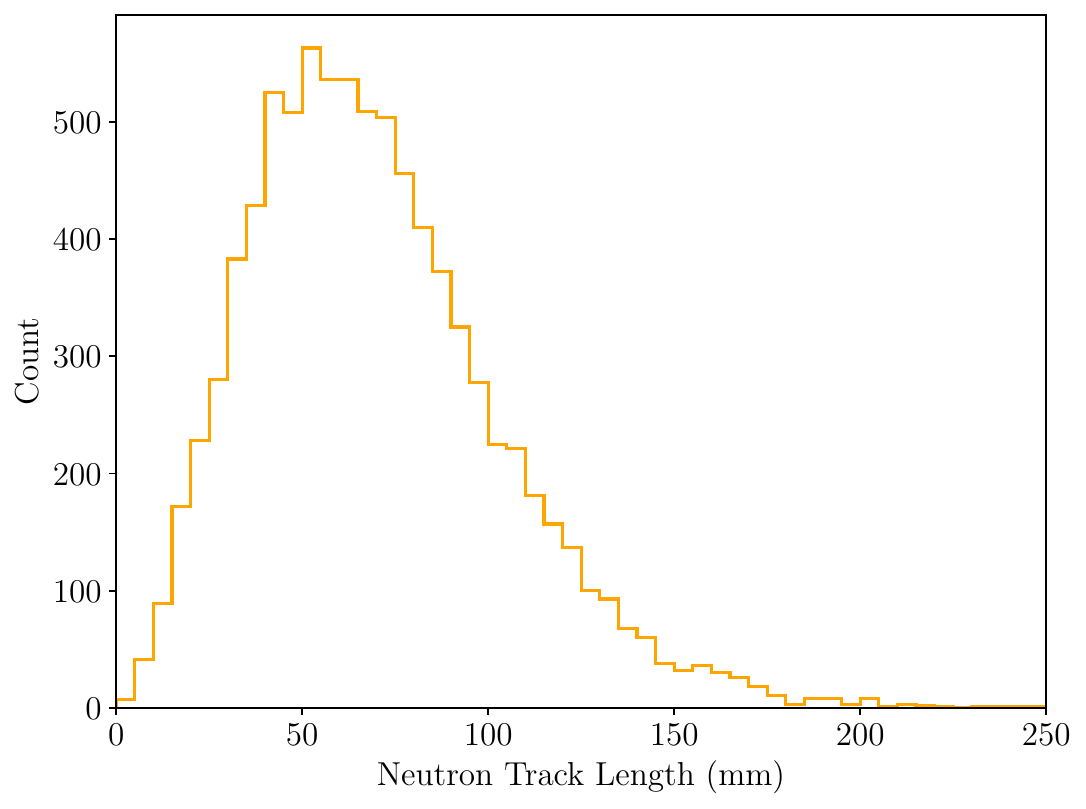}    
    \end{overpic}
    \caption{Histogram of positron track length (top) and neutron track length (bottom) for the 10k event run for 0.1\% ${}^6$Li doped scintillator simulations in RATPAC2.}
    \label{fig:ratpachistograms_positron_neutron_track_length}
\end{figure}

In our simulations, we do not utilize RAT-PAC~2 to its full potential of tracking photons and digitizing PMT signals. Instead, we only use it to simulate IBD positron and neutron pairs in a large, unsegmented homogeneous volume of predefined composition as listed in Table~\ref{tab:li6_loading} and as shown in Fig. \ref{fig_RATPAC_IBD_visual}. 
The IBD positron and neutron are generated according to the reactor antineutrino spectrum implemented in RAT-PAC~2. The corresponding track lengths for the positron and neutron are shown in Fig.~\ref{fig:ratpachistograms_positron_neutron_track_length}. The prompt position is chosen randomly within the segment. Additionally, a grid is overlaid on the simulated monolithic volume to emulate the different segment sizes. To generate binned distribution matrices from different source directions, the grid is rotated, and the distribution of prompt and delayed events is calculated for each angle, as shown in Fig. \ref{fig:trackplot3dratpac_001wt}.
Further information on the validation of the simulations may be found in the appendix.

\begin{figure*}[ht]
\centering
\begin{tabular}{ccc}
\includegraphics[width=0.32\linewidth]{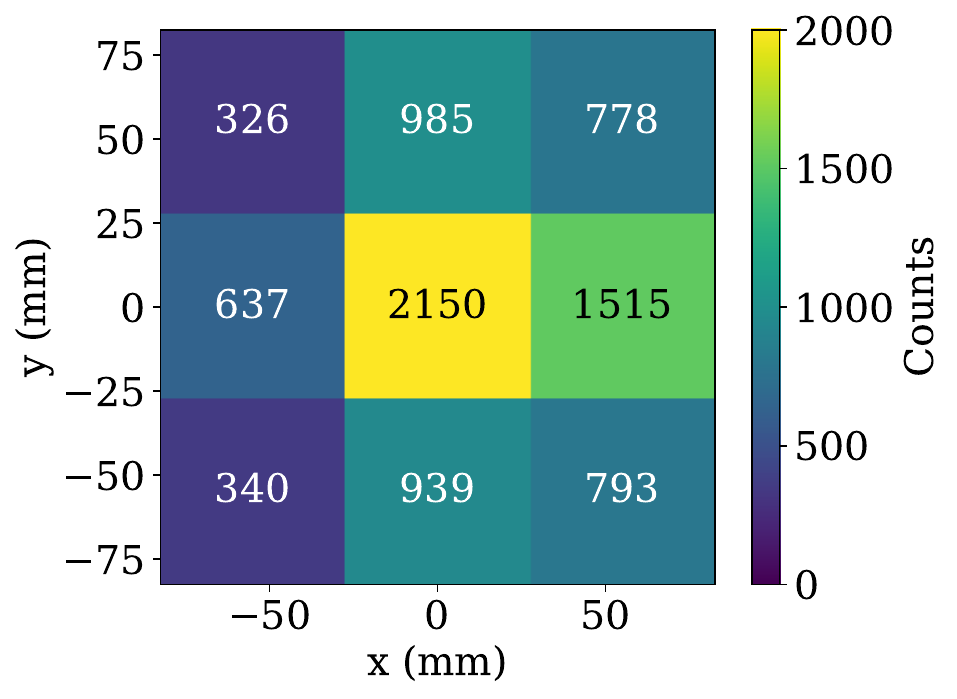}
&
\includegraphics[width=0.32\linewidth]{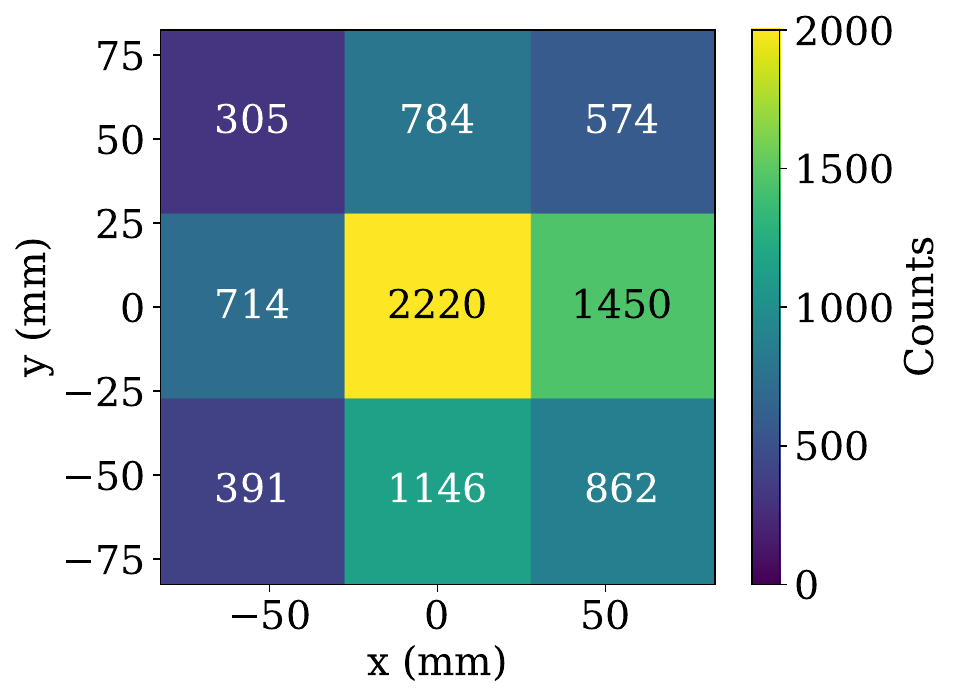}
&
\includegraphics[width=0.32\linewidth]{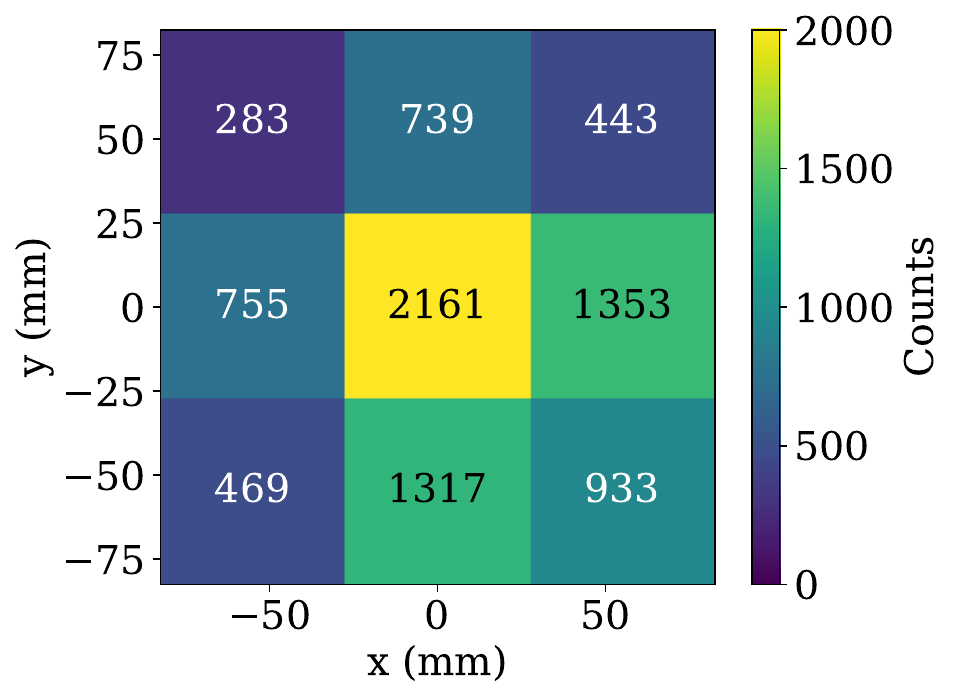} \\
\end{tabular}

    \begin{tabular}{>{\centering\arraybackslash}p{0.32\linewidth} >{\centering\arraybackslash}p{0.32\linewidth}>{\centering\arraybackslash}p{0.32\linewidth}}
        \makebox[0.32\linewidth][c]{%
            \begin{picture}(20,20)
                \put(-5,13){\color{black}\vector(1,0){20}} 
            \end{picture}
        }
        &
        \makebox[0.32\linewidth][c]{%
            \begin{picture}(20,20)
                \put(-3,17){\color{black}\vector(2,-1){18}} 
            \end{picture}
        }
 &
        \makebox[0.32\linewidth][c]{%
            \begin{picture}(20,20)
                \put(-1,19){\color{black}\vector(1,-1){14}} 
            \end{picture}
        }        \\
\hspace{-0.5cm}$\vartheta_\nu = 0^\circ$ & $\vartheta_\nu = -30^\circ$ & $\vartheta_\nu = -45^\circ$ 
    \end{tabular}
    \caption{Examples of binning distributions for incoming angles $\vartheta_\nu=0^\circ,-30^\circ,-45^\circ$, 0.1\%\ ${}^6$Li-doped. The direction of antineutrinos is indicated by the arrows below the plots.}
    \label{fig:trackplot3dratpac_001wt}
\end{figure*}

\section{Developing An Algorithm for Determining Direction and Resolution based on Pattern Matching}
\label{sec:deltaphi}

\begin{figure}[ht!]
        \setlength{\unitlength}{.1\linewidth}
\begin{picture}(10,4)

        \multiput(3,0.5)(.5,0){6}{\color{black}\line(0,1){2.5}} 
        \multiput(3,0.5)(0,0.5){6}{\color{black}\line(1,0){2.5}}

          \thicklines
        \multiput(4., 1.50)(.025,0){20}{\color{black}\line(0,1){0.5}} 
          \thinlines

        \put(4.25,0){\color{black}\vector(0,1){4}} 
        \put(2.5,1.75){\color{black}\vector(1,0){5}} 
        \multiput(4.2,0.25)(0,.5){7}{\color{blue}\line(1,0){0.1}}
        \multiput(2.75,1.7)(.5,0){7}{\color{blue}\line(0,1){0.1}} 

        \put(3.9,3.7){\color{black}$y$}
        \put(7.3,1.4){\color{black}$x$}

        \put(4.25,1.75){\color{blue}\arc[0,45]{2}} 
        \put(2.5,0){\color{blue}\vector(1,1){4}} 
        \put(5.7,3.7){\color{black}$45^\circ$}

        \put(2.5,0.5){\color{blue}\vector(7,5){5}} 
        \put(2.5,1){\color{blue}\vector(7,3){5}} 
        \put(2.5,1.5){\color{blue}\vector(7,1){5}} 

        \put(2.5,1.25){\color{red}\vector(7,2){5}} 
        \put(7.5,2.5){\color{red}$\theta_\nu^{exp}$}
         \put(7.5,2.){\color{blue}$\theta_\nu^{MC_1}$}
        \put(7.5,3){\color{blue}$\theta_\nu^{MC_2}$}
        \put(7.5,3.7){\color{blue}$\theta_\nu^{MC_3}$}

\end{picture}
\caption{A diagram explaining the idea behind a pattern-matching algorithm for IBD segmented detectors. The experimental/empirical data set (with {\it a priori} unknown neutrino angle $\theta_\nu^{exp}$) is compared against theoretical/simulation data sets (of known neutrino angles $\theta_\nu^{MC_i}$). Due to the symmetry and square geometry unique distributions only obtained between 0$^\circ$ and 45$^\circ$ angles. MC$_1$, MC$_2$, and MC$_3$ denote Monte Carlo data sets corresponding to that particular neutrino angle. The 5$\times$5 grid is shown for illustrative purposes.}
\label{fig_algo_skewers}
\end{figure}
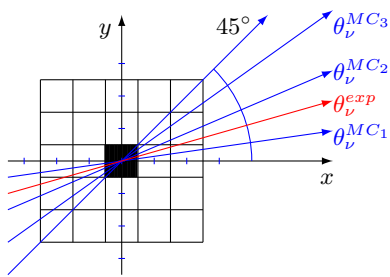

The algorithm was developed contemporaneously with this work and is described in a separate paper published by our research group~\cite{10.1063/5.0315079}. 
The algorithm uses the Frobenius norm of the difference between a simulated or known dataset and the unknown dataset. 
The unknown dataset is the matrix holding the neutron captures collected from a segmented detector, as described above. 
For our application to neutrino directionality, a reference dataset is created by simulating IBD events in a plastic scintillating detector. 
We use the spatial coordinates of the simulated neutron capture event locations, rotate them about the origin or the ``prompt'' segment of the detector geometry, and bin them into a 2D histogram. 
We treat the histogram data as a matrix array. 
At each rotation, we compare the simulated reference matrix dataset with the unknown matrix dataset via the Frobenius norm of the difference between the two matrices. 
The rotated dataset that is closest to the unknown dataset produces a minimum which we find by fitting the Frobenius norm with an absolute value sine function $|\sin \vartheta|$ (the mathematical proof is provided in our original study~\cite{10.1063/5.0315079}).
The angle that is the location of the minimum of this fit is treated as the best reconstructed direction from the algorithm.

We begin with generating an array centered at the truth IBD vertex with a bin (position in the array) correlating to a segment displacement from the vertex. The segment sizes are analyzing are generated to correlate with the 3  segmented detector geometries of interest: SANDD, MAD,  and Prospect. Since these are real detectors with real data we can compare the real cloud with the simulated cloud. Below is the methodology to do the comparison using the $L^2$ norm of the difference of two sets, where $i$ is the index for the number of iterations:
\begin{enumerate}
    \item Set the segment size of generated data to that of the detector you are comparing against. This changes the binning resolution and is necessary to capture the same processes between ``artificial data'' (data points produced from RATPAC2) and data collected from a detector.
    \item At each rotational angle with respect to the neutrino wind, generate the same number of usable events to construct a neutron capture cloud that is then binned in the array from IBD vertex to final capture, matching the number of bins used to avoid addition complications. 
    
    $\hat{R_{\theta}}[$SimulatedSet$_i] = [$SimulatedSet$_i](\theta) $
    \item Subtract each generated array from the array of real data from the detector.

     $[$SimulatedSet$_i](\theta) - [$TrueSet$] = [$DifferenceSet$_i](\theta)$
    \item Take the $L^2$ norm of each difference array. 
    
    $L^2_i(\theta) = \sqrt{\sum_{kj}({x_{kj})_i^2}} $\\$ (x_{kj})_i \in [$DifferenceSet$_i](\theta)$.
    
    \item The angle $-\pi \leq\vartheta_i\leq\pi$ that corresponds to the lowest value of the fit function applied over the $L_i^2(\theta)$, is the calculated angle that the neutrino wind is coming from, with respect to the orientation of your neutrino detector. Normalized
    
    $\min_i\{|\sin{(\theta)}|\} = |\sin{(\vartheta_i)}|$

\end{enumerate}

The steps above are the exact steps of the directionality algorithm \cite{10.1063/5.0315079}, schematically shown in Fig.~\ref{fig_algo_skewers}. This is what we call an iteration. As a result of a single iteration --- the FND (Frobenius norm of the difference matrix, also known as L$^2$ norm.) is calculated for each angle, and an example of the algorithm is shown in the Appendix in Fig.~\ref{fig:CFND_1iteration} for $n=10,100,1000$ events. (for number of events equal 1,000 --- not to be confused with the number of iterations as explained below).

\subsection{Approximating error on the angular uncertainty}

We then run $\lambda$ iterations by repeating steps 1--5 described above, essentially applying this algorithm multiple times to obtain an angular spread of the angular uncertainty. To obtain a measure of the angular uncertainty we run the following final steps:
\begin{enumerate}

    \item Iterate this process over $i$ such that the following statistical analysis is valid ($\lambda \geq 30$)
    \item The reported direction of the incoming neutrino wind is calculated by taking the average of all determined $\vartheta_i$
\[\bar\vartheta=\sum^\lambda_i\frac{\vartheta_i}{\lambda} \]
    
    \item The distribution function of $[\vartheta_i]$ over a circle can be described as a circular distribution, we report the error in this method as the circular standard deviation of the distribution. A derivation for the circular standard deviation as well as the uncertainty of uncertainty is presented in the Appendix.
\end{enumerate}

\begin{figure}[ht]
    \includegraphics[width=1\linewidth]{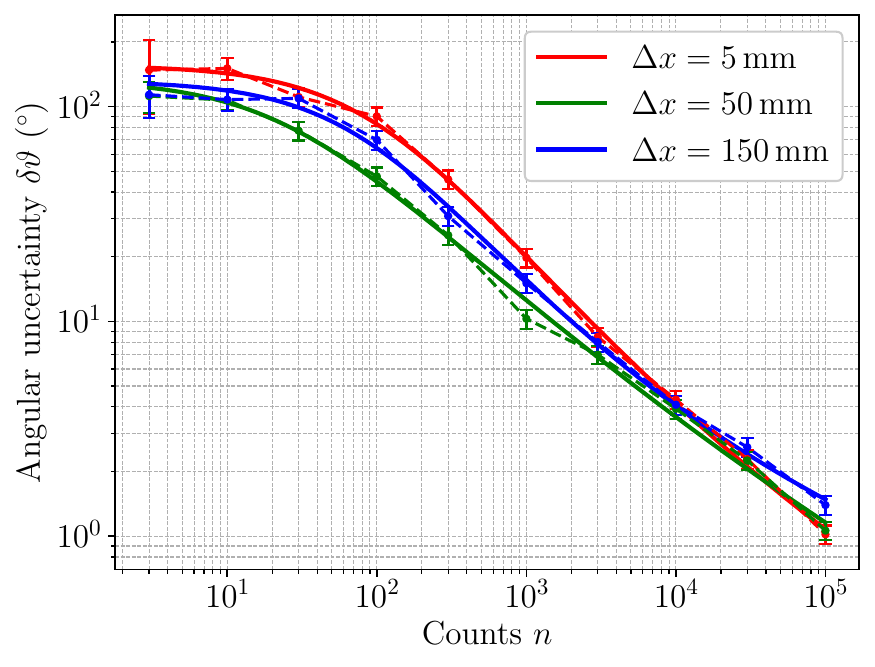}
    \caption{Angular uncertainty plot comparing the performance of three different square segment sizes: 5 mm, 50 mm and 150 mm. Notice that the smallest segment size begins to perform better than the other two for higher counts. Smaller segment sizes outperform larger ones for higher counts simply due to superior resolution.}
    \label{fig:POI_segsize_uncert}
\end{figure}

After performing these calculations for a specific number of IBD events, an angular uncertainty is calculated. Therefore we repeat this for a fixed number of events. The resulting plot of angular uncertainty (with corresponding error bars) is shown in Fig.~\ref{fig:POI_segsize_uncert}.

The fit function used for the angular uncertainty distributions in Fig.~\ref{fig:POI_segsize_uncert}, for which the associated parameters for each fit are given in Table~\ref{tab:fit_parameters}, was
\begin{align}
y_\mathrm{fit}(n)=\exp\left[\left(a-d\tan^{-1}{\left(\frac{\log{(n+b)}}{c}\right)}\right)\ln 10\right].
\end{align}

\begin{table}[h]
\centering
\begin{tabular}{c c c c}
\hline
Parameter & 5 mm & 50 mm & 150 mm \\
\hline
$a$ & $4.225 \pm 0.821$ 
    & $3.005 \pm 0.283$ 
    & $4.718 \pm 1.682$ \\

$b$ & $98.66 \pm 49.79$ 
    & $21.69 \pm 12.73$ 
    & $91.38 \pm 51.85$ \\

$c$ & $4.941 \pm 2.382$ 
    & $7.551 \pm 3.975$ 
    & $2.866 \pm 1.557$ \\

$d$ & $5.298 \pm 0.652$ 
    & $5.030 \pm 1.670$ 
    & $4.329 \pm 0.634$ \\
\hline
\end{tabular}
\caption{Fit parameters for 5 mm, 50 mm, and 150 mm configurations.}
\label{tab:fit_parameters}
\end{table}

\begin{figure}[ht]
    \includegraphics[width=1\linewidth]{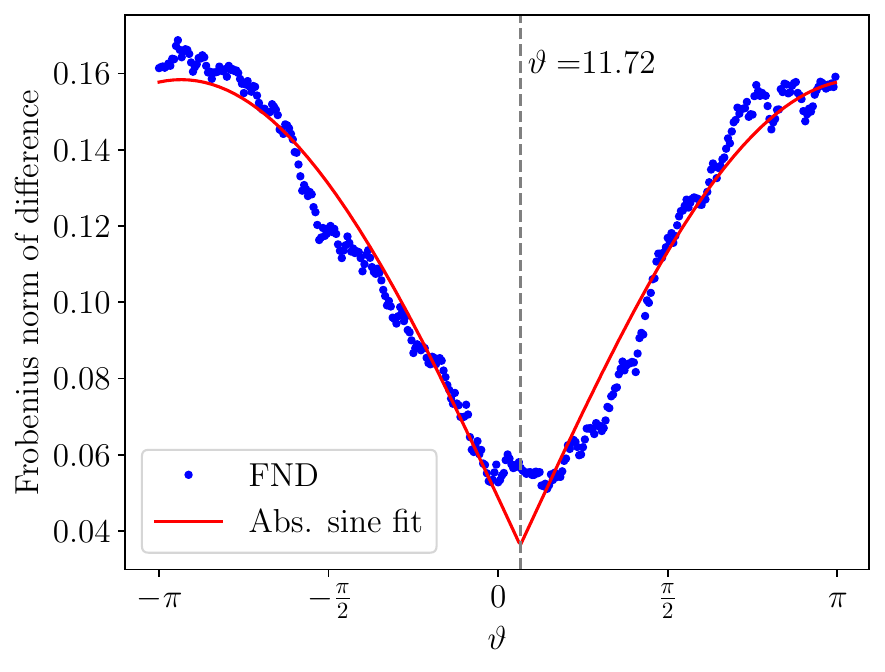}
    \caption{Frobenius norm of the difference between an artificial reference dataset produced for a true angle of $\vartheta_{true} = 0^\circ$ and the datasets for all angles within $\pm\pi$ radians from either side of the reference angle. Uses $50 \,\text{mm}$ segment size and $n=1000$. The FND of the neutron capture distribution data was fit with an absolute value of sine function as prescribed by the theoretical proof for the direction algorithm \cite{10.1063/5.0315079}.}
    \label{fig:CFND_1iteration}
\end{figure}

\begin{figure}[ht]
    \includegraphics[width=1\linewidth]{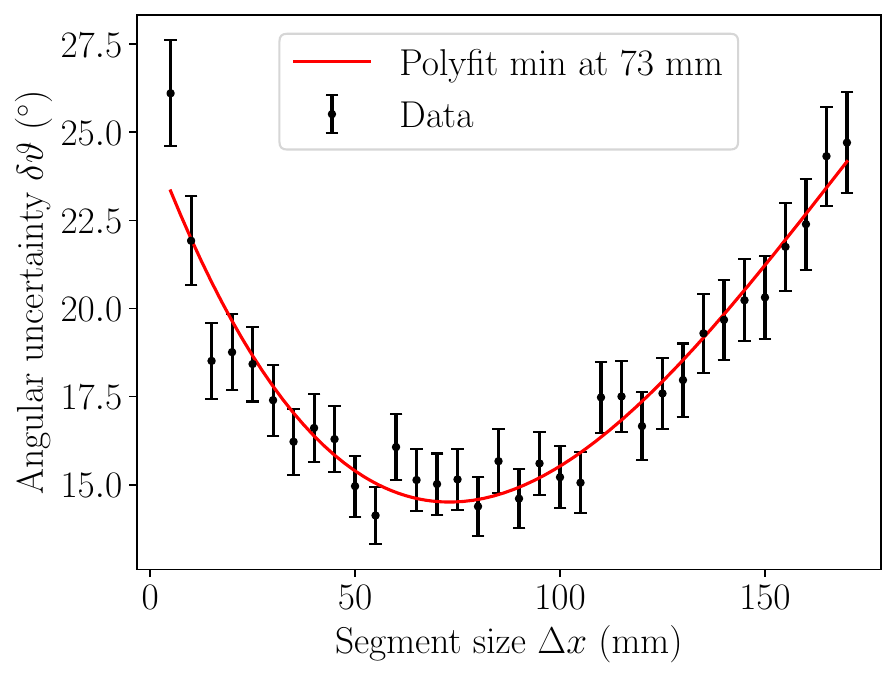}
    \caption{Optimal segment size plot with a fitted minimum (ideal segment size in the low $n$ limit) of 73 mm. Each point was run 300 iterations at $n=300$. The average track length of the neutron (Euclidean distance from IBD vertex to capture location) calculated from the $10^6$ event fiducial dataset came out to be roughly 70 mm. The ratio of the minimum of this fit to the average track length is $73/70=1.04$, which shows that the ideal segment size is roughly the same as the average track length of the neutron before capture---an interesting result. There are two effects going on in the low $n$ limit: (1) when the segment size is very small, the binned distribution of neutron captures is sparse; (2) when the segment size is very large, more events fall into the center element of the binned distribution which carries no directional information upon rotation. These two behaviors result in the $\Delta x=50$~mm segment having the best angular resolution of the three chosen segment sizes.}
    \label{fig:sweet_spot}
\end{figure}

\section{Results}
Our primary plot of interest  (POI) Fig.~\ref{fig:POI_segsize_uncert} shows the relationship between  different segment sizes, IBD events, and directional uncertainty. The merit of our primary finding is a critical improvement in the quantification of directionality. As our discussion of \cite{Duvall:2024cae} and Fig.~\ref{fig_old_method} above highlights, we have identified the need for a statistical method that is mathematically rigorous at low event count. 

The additional merit of our study is publishing on how projects may now find and design projects around the neutron capture ``sweet spot'' (for a given $^6$Li dopant level). Our second POI is Fig.~\ref{fig:POI_usable_event_segsize_uncert} which shows the relationship for usable events, where usable events are counted as those that lie outside the range of the center segment of the detector. In Fig.~\ref{fig:POI_segsize_uncert}, note that the 5-mm uncertainty is larger than the 150-mm. This is due to sparse matrices that are dependent on the neutron mean-travel distance before capture, see Fig.~\ref{fig_segmentation_scaling}.  Via our results from the FND (see Fig.~\ref{fig:POI_segsize_uncert}, due to the large spread of the neutron distribution there is high angular uncertainty for low $n$) we can accurately quantify uncertainty in the low $n$ regime by running multiple iterations of the algorithm, showing convergence to a singular angular uncertainty. With sufficient samples and at suitable detector size, the FND converges to a function that is proportional to $|\sin((\vartheta_0-\vartheta)/2)|$, a detailed derivation can be found in \cite{10.1063/5.0315079}, what we refer to as the ``v-shaped" plots, as shown in Fig.~\ref{fig:CFND_1iteration}.

\section{Conclusion}

We present a directionality algorithm based on pattern-matching. This converges on a higher (and thus more realistic) angular uncertainty at low number of events. 
In our previous study \cite{Duvall:2024cae}, we analyzed different detector geometries via the \textit{Chooz} method and showed that there are statistical limits to directionality. 
The previous method superficially creates a notion of angular resolution of $\mathcal{O}(10)$ degrees at tens of events due to its analytical formulation.
This method clearly identifies the inflection point ($n=30$ in Fig.~\ref{fig:POI_usable_event_segsize_uncert}) below which the notion of resolution becomes artificial. 
 
The algorithm we present requires a set of data from simulation to provide the pattern to match to determine the source direction using data from experiment. In order to facilitate an accurate directional reconstruction, the simulation data should represent  the detector used for the experimental data. While the algorithm was tested using two dimensionally segmented detector geometries, in principle it is possible to apply it to three dimensionally segmented geometries as well.

We have considered some potential applications of this algorithm, including nuclear safeguards and geo-neutrino studies. First, are small modular reactors, where one has multiple cores. Second, in situations with a low number of events or long standoff distances. Follow-on studies will examine the algorithm's performance in the presence of various backgrounds. Planned future studies will investigate incorporating various statistical methods in directional analysis, applying this method to detectors of different geometry such as monolithic and separated, as well as with different detection media such as water Cherenkov, and applying the method to three dimensionally segmented detectors. Applying this method to different deployment scenarios, such as small modular reactors, will investigate some of the underlying assumptions, such as neglecting \(\frac{1}{r^2}\) dependence and non-zero angular size of nearby reactors are all being considered for future work.

\section{Acknowledgments}
We thank Marc Bergevin for his help with RAT-PAC~2, as well as Livermore High-Performance Computing. We also thank Mark Duvall, whose insights into the directionality problem in segmented detectors led us to do this work.
This work was partially supported by the Consortium for Monitoring, Technology, and Verification, and the University of Hawai`i HEP grant DE-SC0010504.

This work was performed under the auspices of the U.S. Department of Energy by Lawrence Livermore National Laboratory under Contract DE-AC52-07NA27344. LLNL-JRNL-2016175.

\section{Data Availability Statement}

The data that support the findings of this study were generated by numerical simulations. The source code and parameters used to generate the simulations is publicly available~\cite{mtv-directionality}. The code used to analyze the data is available from Ref.~\cite{neutrino-directionality}.

\appendix*
\section{Appendix}

\subsection{Appendix A: Uncertainty of Uncertainty and Circular Statistics}\label{app:uncert_of_uncert_and_circularstats}
In order to validate our algorithm method, it was essential to quantify error on this uncertainty measurement and compare with the method reported in \cite{Duvall_2022}. To quantify angular uncertainty of the detector we produce a circular distribution of $\lambda$ angles  $\vartheta_i$, where $i\in[1,\lambda]$. Each of these angles is produced by a unique run of the direction algorithm---as described in our paper \cite{10.1063/5.0315079}, of which each run has a sample size $n$ of the neutron capture location distribution. We then use the SciPy von Mises MLE fit to compute the mean and angular uncertainty of the circular distribution, which uses the following well-known circular statistics. We conveniently represent each angular data point as a number on the complex plane, expressed as
\begin{align}\label{eq:zdef}
z_i\equiv e^{i\vartheta_i}=\cos\vartheta_i+i\sin\vartheta_i,
\end{align}
where $i$ is the imaginary number $i=\sqrt{-1}$ and is not to be confused with the number of iterations subscript  used to denote the angle $\vartheta_i$.
As prescribed in \cite{Fisher_1993}, the mean resultant vector is defined as
\begin{align}
\bar \rho\equiv\frac{1}{\lambda}\sum_{i=1}^\lambda z_i.
\end{align}
It becomes convenient to define the mean sine and cosine as
\begin{subequations}
\begin{align}
\overline S &\equiv \frac{1}{\lambda}\sum_{i=1}^\lambda \sin\vartheta_i
\\
\overline{C} &\equiv \frac{1}{\lambda}\sum_{i=1}^\lambda \cos\vartheta_i.
\end{align}
\end{subequations}
This allows us to write the mean resultant vector as
\begin{align}
\overline \rho = \overline C + i\overline S.
\end{align}
The sample mean is given by the complex argument
\begin{align}
\overline \vartheta = \text{Arg}\,(\overline \rho)=\text{Arg}\,(\overline C + i\overline S)=\text{atan2}\,(\overline S, \overline C).
\end{align}
where atan2 is the tuple-function equivalent of $\text{tan}^{-1}\, {y/x} $.
The mean resultant length is the normal of the mean resultant vector
\begin{align}
\overline R=|\overline \rho|=\left|\overline C + i\overline S\right|=\sqrt{\overline S^2 + \overline C^2}.
\end{align}
The circular standard deviation is given in terms of the mean resultant length as
\begin{align}
\sigma=\sqrt{-2 \ln\overline R},
\end{align}
which we define as the angular uncertainty $\delta \vartheta\equiv\sigma$ of a single measurement. 
Therefore, we report the final value as $\overline\vartheta\pm\delta\vartheta$. As a final note, we estimate the uncertainty on the uncertainty as
\begin{align}
\delta(\delta \vartheta)=\delta^2\vartheta=\frac{\sigma}{\sqrt{\lambda}}.
\end{align}
This quantity was used to calculate the error bars in the angular uncertainty plots in Figs.~\ref{fig:POI_segsize_uncert} and~\ref{fig:POI_usable_event_segsize_uncert}.

Consider a scenario where we want to simulate many dart players and using circular statistics, infer what angle they threw from. Let us suppose our simulation software is limited to only simulating a direct beam from $\theta = \phi = 0$. Rather than simulate many positions of players however, the board itself can be rotated about, up to an oblique angle $\pi$. 
Stepping out of analogy, your detector volume may be rotated (or translated) in the world-volume but the neutrino-flux angle is not a beam parameter.

Consider a second scenario: you are hiding Easter eggs for children. They are durable eggs and you throw them in a straight line but they bounce around a bit depending on their kinetic energy upon scattering with the mix of open grass, shrubs, and trees in front of you; such that eggs will randomly experience elastic and inelastic scattering. Some of the eggs get caught in bushes. Because you know a priori the location of all eggs, to test if an algorithm quantifies directionality, you can close your eyes and toss out a tick-tack-toe grid (think of the 2d projection of segmented detectors which are rectangular prisms). The Frobenius norm method described in this paper quantifies how much one needs to rotate that grid back into a $\theta = 0$ position.

Therefore, if there is any statistically significant difference in the distribution between those \textit{after} aligning the grids, we know there is a true angular difference between the ``measured'' source and true $\theta = \phi = 0$, a.k.a. a difference only possible if you accidentally threw the eggs at a non-zero angle. 

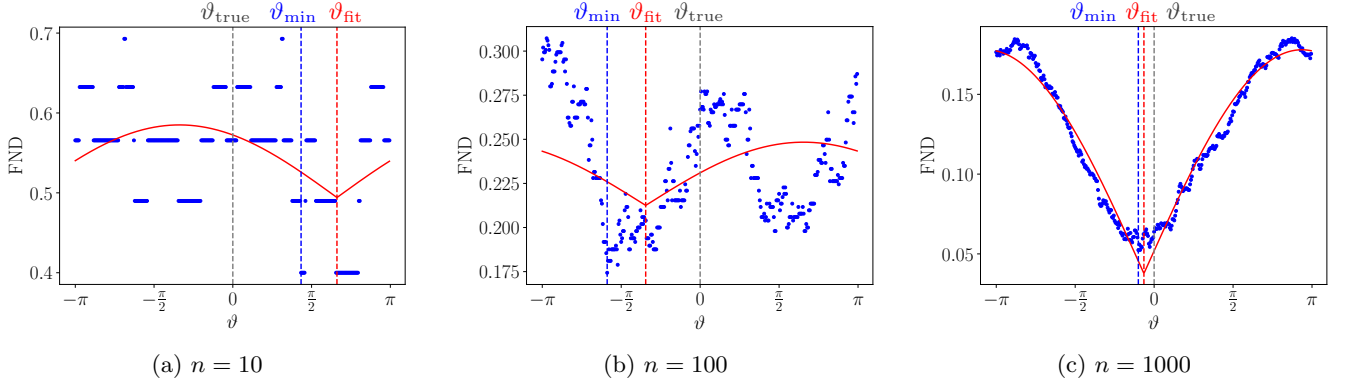
\begin{figure*}[htbp]
\centering

\begin{minipage}{0.32\textwidth}
    \centering
    \begin{tikzpicture}
        \node[inner sep=0]
            {\includegraphics[height=1.675in]{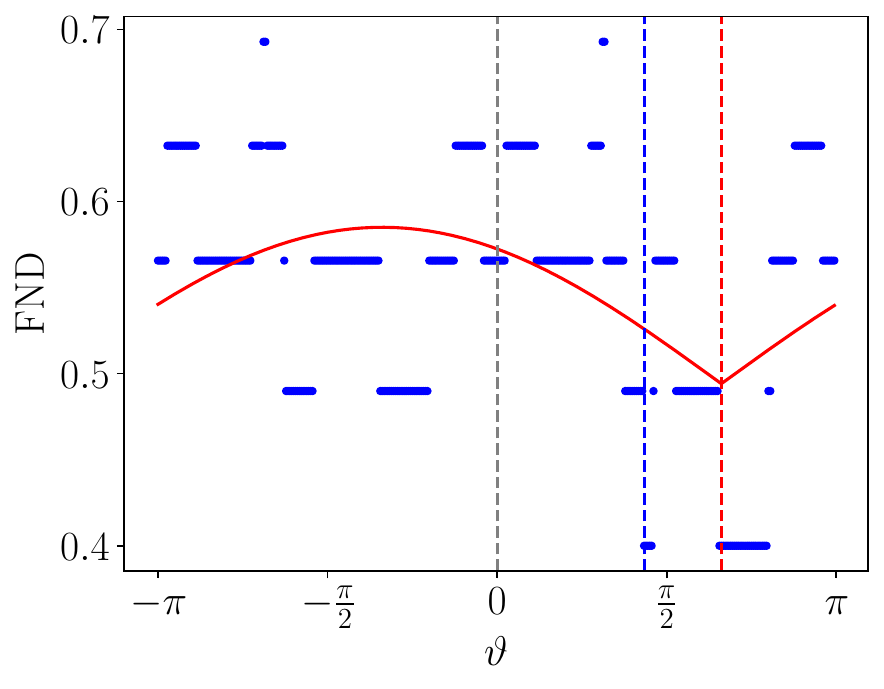}};
        
        \node[red, font=\small] at (1.85,2.225) {$\vartheta_\mathrm{fit}$};
        \node[blue, font=\small] at (1.15,2.225) {$\vartheta_\mathrm{min}$};
        \node[gray!50!black, font=\small] at (0.25,2.225) {$\vartheta_\mathrm{true}$};
    \end{tikzpicture}

    \vspace{2pt}
    (a) $n=10$
\end{minipage}
\hfill
\begin{minipage}{0.32\textwidth}
    \centering
    \begin{tikzpicture}
        \node[inner sep=0]
            {\includegraphics[height=1.675in]{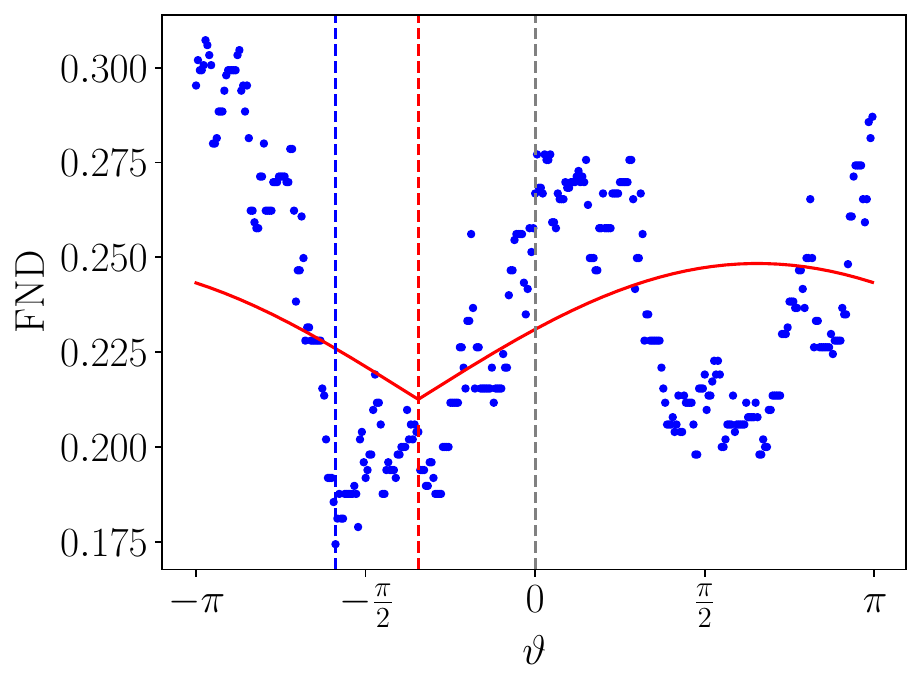}};
        
        \node[red, font=\small] at (-0.25,2.23) {$\vartheta_\mathrm{fit}$};
        \node[blue, font=\small] at (-0.9,2.23) {$\vartheta_\mathrm{min}$};
        \node[gray!50!black, font=\small] at (0.45,2.23) {$\vartheta_\mathrm{true}$};
    \end{tikzpicture}

    \vspace{2pt}
    (b) $n=100$
\end{minipage}
\hfill
\begin{minipage}{0.32\textwidth}
    \centering
    \begin{tikzpicture}
        \node[inner sep=0]
            {\includegraphics[height=1.675in]{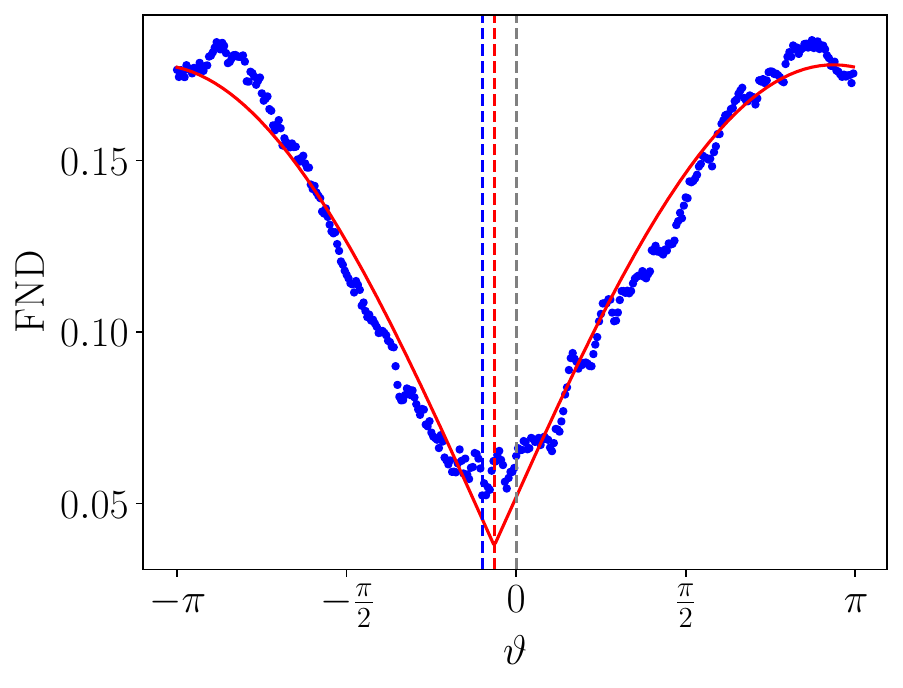}};
        
        \node[red, font=\small] at (0.25,2.23) {$\vartheta_\mathrm{fit}$};
        \node[blue, font=\small] at (-0.4,2.23) {$\vartheta_\mathrm{min}$};
        \node[gray!50!black, font=\small] at (0.9,2.23) {$\vartheta_\mathrm{true}$};
    \end{tikzpicture}

    \vspace{2pt}
    (c) $n=1000$
\end{minipage}

\caption{Explanation of angles and spread. The FND here is calculated from the RATPAC2 neutron capture data with a segment size of 50 mm. The absolute value of sine fit is applied as shown.}
\label{fig_spread}
\end{figure*}

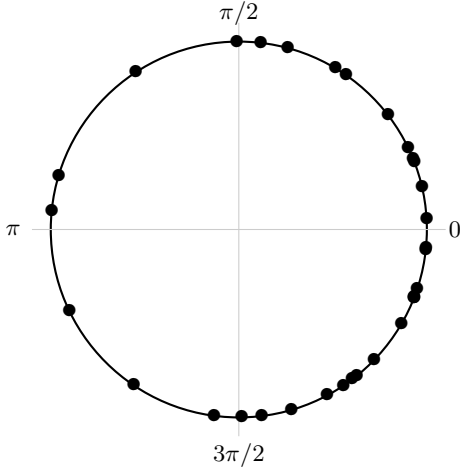
\begin{figure}[htbp]
\centering
\begin{tikzpicture}[scale=2.5]

\def\r{1}

\draw[thick] (0,0) circle (\r);

\draw[gray!40] (-1.1,0) -- (1.1,0);
\draw[gray!40] (0,-1.1) -- (0,1.1);

\node at (1.15,0) {$0$};
\node at (0,1.15) {$\pi/2$};
\node at (-1.2,0) {$\pi$};
\node at (0,-1.2) {$3\pi/2$};

\foreach \angle in {21.08390459, -73.77888731, 163.36116687, -97.55230599, 174.34410792, 13.10232092, 55.18831205, 22.117807, -30.09141061, -61.95640066, -154.24266907, 3.0804383, 59.09075331, -89.09277446, -51.19303388, -82.98397561, 37.41648349, 123.22824185, -21.36105465, 25.73459212, -43.9460415, -21.261096, -52.92347806, 74.90147315, -123.91736139, -5.62295241, -21.15689251, -6.216944, 83.28095805, -56.24029139, 90.63476496, -18.49481349}
{
    \node [font=\large] at ({\r*cos(\angle)}, {\r*sin(\angle)}) {$\bullet$};
}

\end{tikzpicture}
\caption{Uncertainty of uncertainty. Each point is generated by an iteration of the direction algorithm, i.e. $\vartheta_\mathrm{fit}$. In the darts simulation analogy, each iteration is simulating many players all with a fixed number of darts (see Fig.~\ref{fig:POI_usable_event_segsize_uncert})}
\label{fig_spread2}
\end{figure}

To further clarify the algorithm and the uncertainty calculation, we include a few figures to this appendix, Fig.~\ref{fig_spread} shows the behaviour of the CFND V-plots at various number of IBD events (within a single iterations).
For a fixed number of IBD events, we show how the angle is inferred for various number of iterations, shown in Fig.~\ref{fig_spread2}.

\subsubsection{Use of synthetic data}
For low counts---most notably $n=10$ and below---some iterations of the algorithm return no directional information at all. To account for this, synthetic data was inserted where we could not determine directionality, i.e. when the FND data is a constant function. Each point on the uncertainty plot is generated by $\lambda$ iterations of the direction algorithm. Now suppose $\lambda_\mathrm{synthetic}$ is the number out of those $\lambda$ iterations that are synthetic random samples from the range $[-\pi,\pi]$ (for which the FND fit fails due to the data being flat or having no directional information). 
In other words, for such trials, we have $0 \leq \lambda_\mathrm{synthetic} \leq \lambda$. The error on error bars on the angular uncertainty plots are then corrected by 
\begin{align}
\delta(\delta\vartheta)\approx\frac{\delta\vartheta}{\sqrt{\lambda-\lambda_\mathrm{synthetic}}},
\end{align}
where $\delta\vartheta=\sigma$.

\subsection{Appendix B: Usable events}
Below in Fig.~\ref{fig:POI_usable_event_segsize_uncert} is the angular uncertainty plot for usable events. Usable events are counted as those that lie outside the range of the center segment of the detector where the prompt event was recorded.
\begin{figure}[!h]
    \includegraphics[width=1\linewidth]{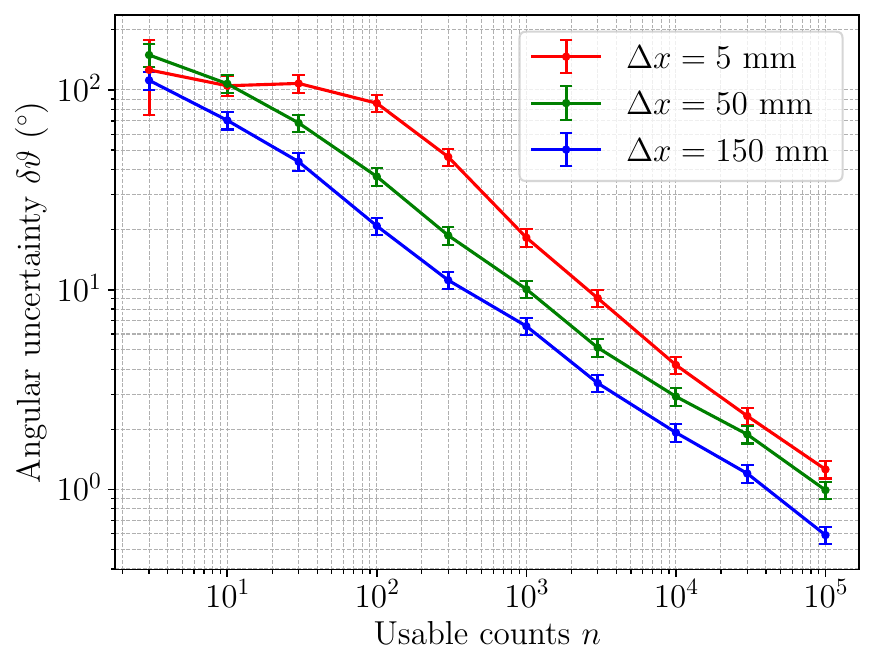}
    \caption{Angular uncertainty plot for usable events.}
    \label{fig:POI_usable_event_segsize_uncert}
\end{figure}

\subsection{Appendix C: simulation validation and parameter space}
\label{app:phys_validation}
\begin{figure*}
    \centering
    \begin{overpic}[width=0.46\linewidth]
    {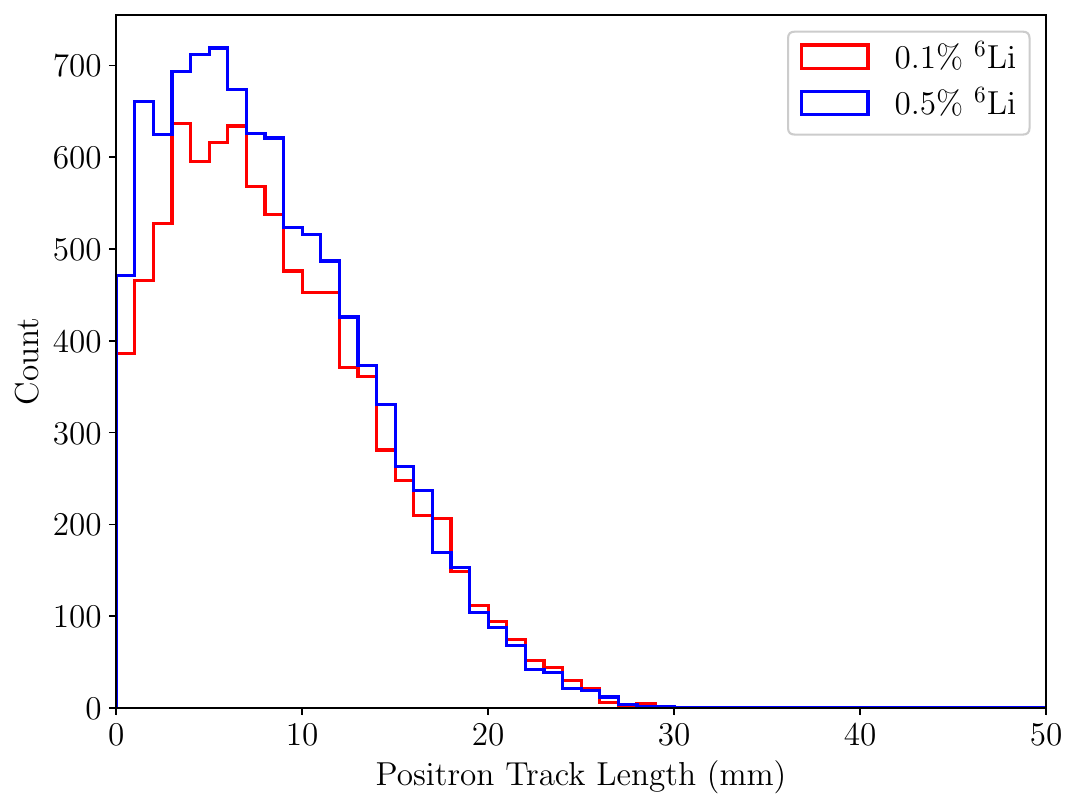}
    \end{overpic}
    \begin{overpic}[width=0.46\linewidth]{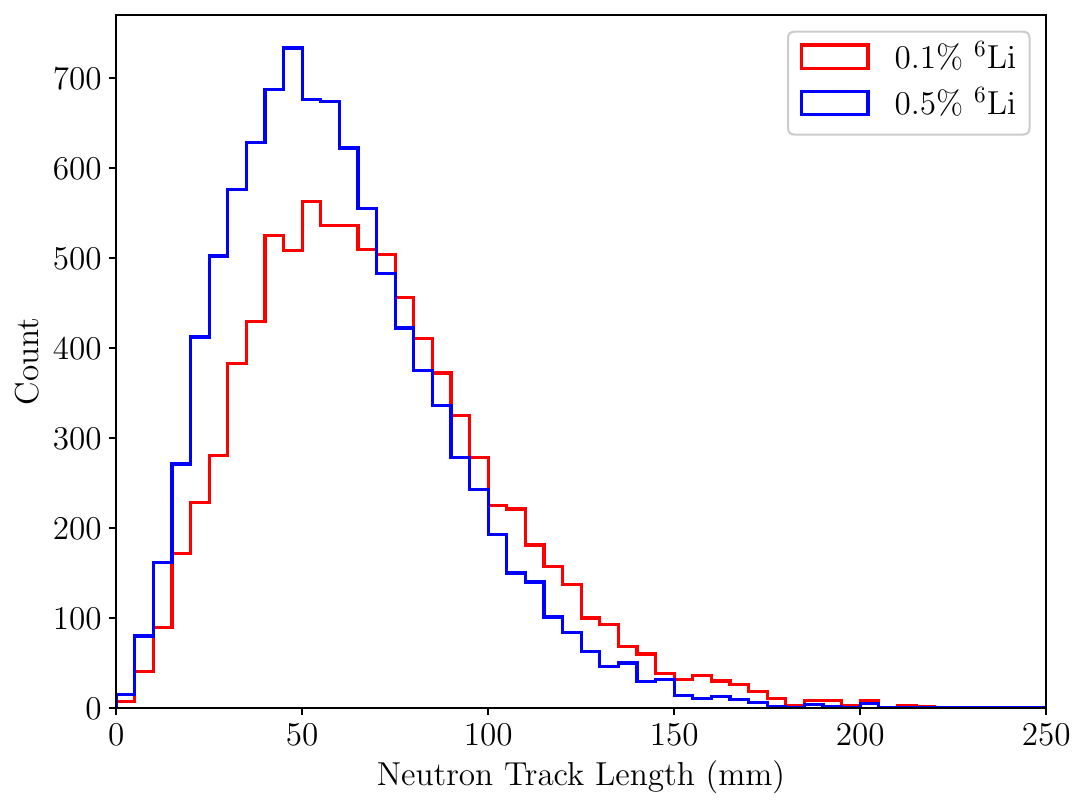}
    \end{overpic}
    \caption{Histogram of positron track length (left) and neutron track length (right) extracted from the 10k event runs for 0.1\% and 0.5\% ${}^6$Li doped scintillator simulations in RATPAC2. In the $0.1\%~{}^6\text{Li}$ doped scintillator, positrons traveled an average distance of $9\text{~mm}$ before annihilation, with a standard deviation of $18\text{~mm}$. Neutrons traveled an average distance of $70\text{~mm}$ with a standard deviation of $34\text{~mm}$.}
    \label{fig:ratpachistograms}
\end{figure*}

\begin{figure*}[ht]
    \centering
    \begin{overpic}[width=0.32\linewidth]
    {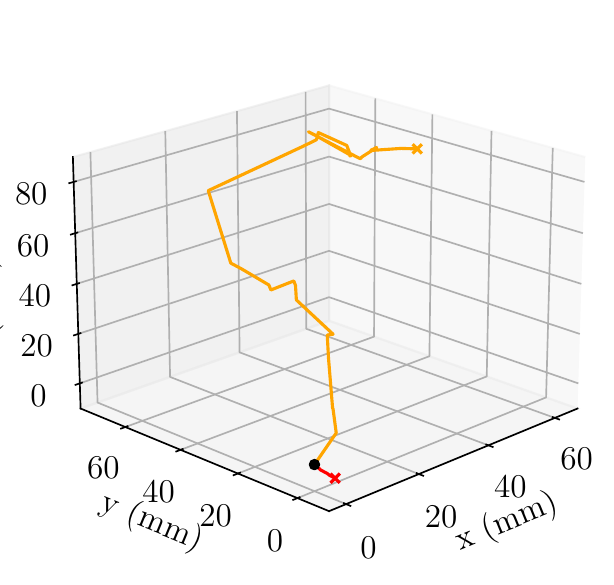}
    \end{overpic}
    \begin{overpic}[width=0.32\linewidth]{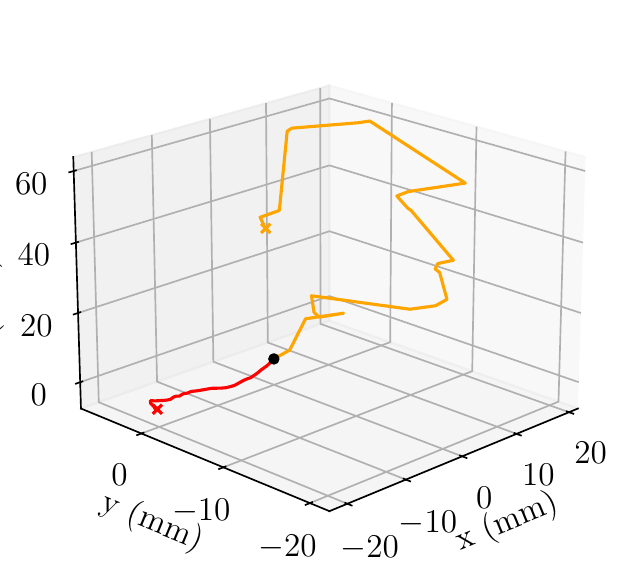}
    \end{overpic}
    \begin{overpic}[width=0.32\linewidth]{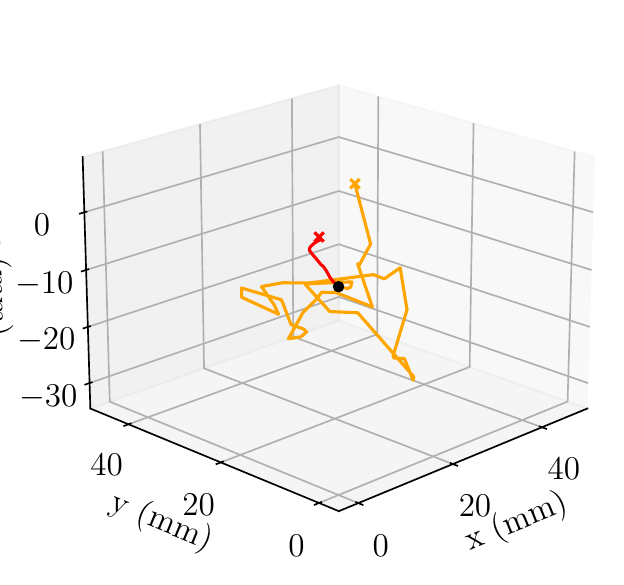}
    \end{overpic}
    \caption{Examples of simulated 3d track plots of neutron and positron tracks. The positron track is shown in red and the neutron track in orange. The positron annihilation location is marked by the red cross and the neutron capture is marked by the orange cross. The true IBD vertex is marked by the black point.}
    \label{fig:trackplot3dratpac}
\end{figure*}

\begin{figure*}[ht]
    \centering
    \begin{overpic}[width=0.31\linewidth]
    {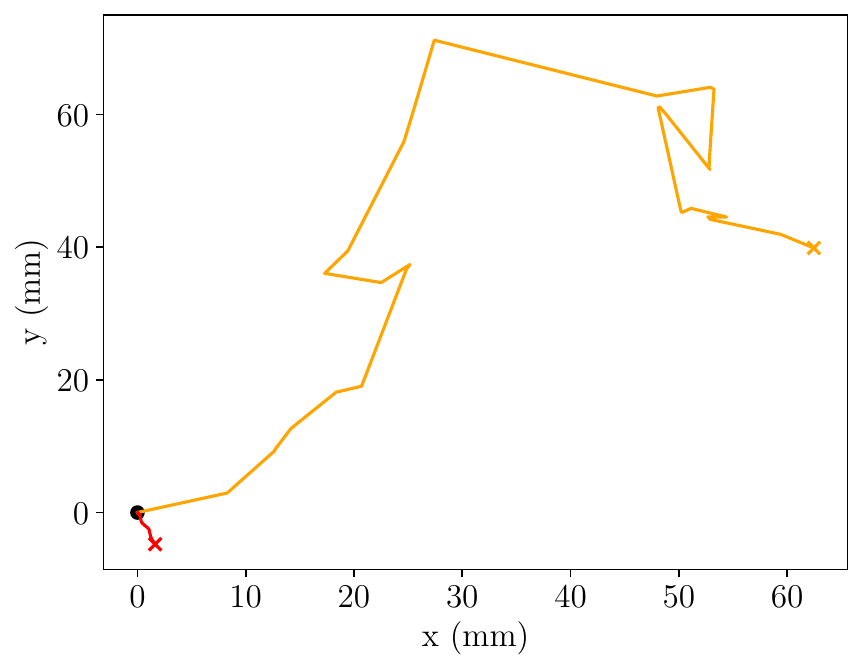}
    \end{overpic}
    \begin{overpic}[width=0.32\linewidth]{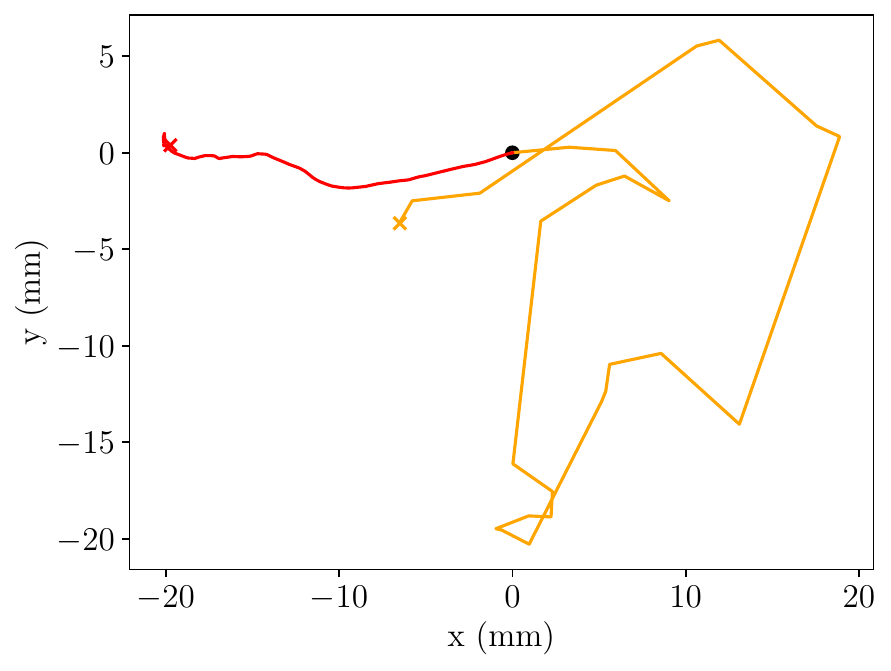}
    \end{overpic}
    \begin{overpic}[width=0.31\linewidth]{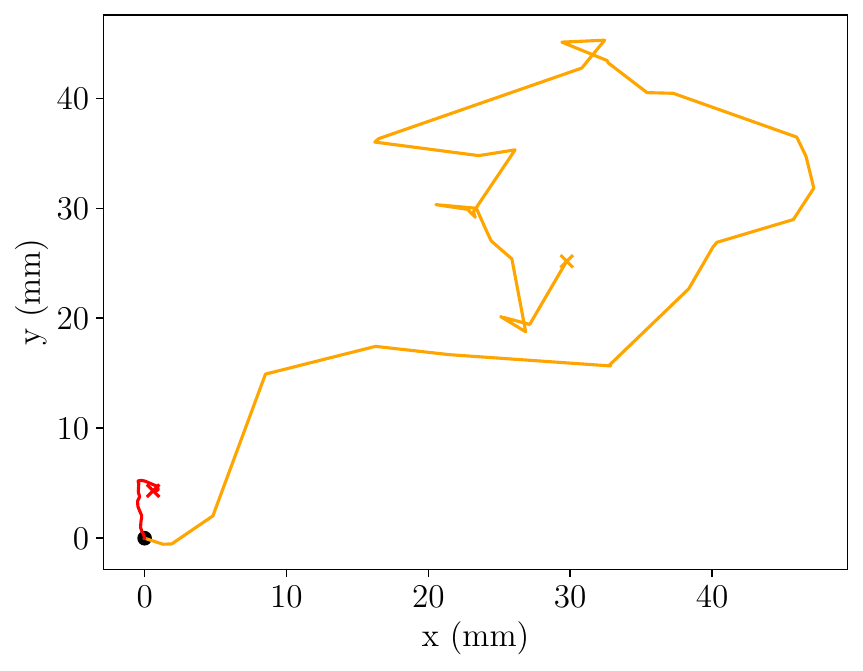}
    \end{overpic}
    \caption{Rotated top view of 3d track plots shown in Fig.~\ref{fig:trackplot3dratpac}.}
    \label{fig:trackplot3dratpactopview}
\end{figure*}

\begin{figure}[ht]
    \includegraphics[width=1\linewidth]{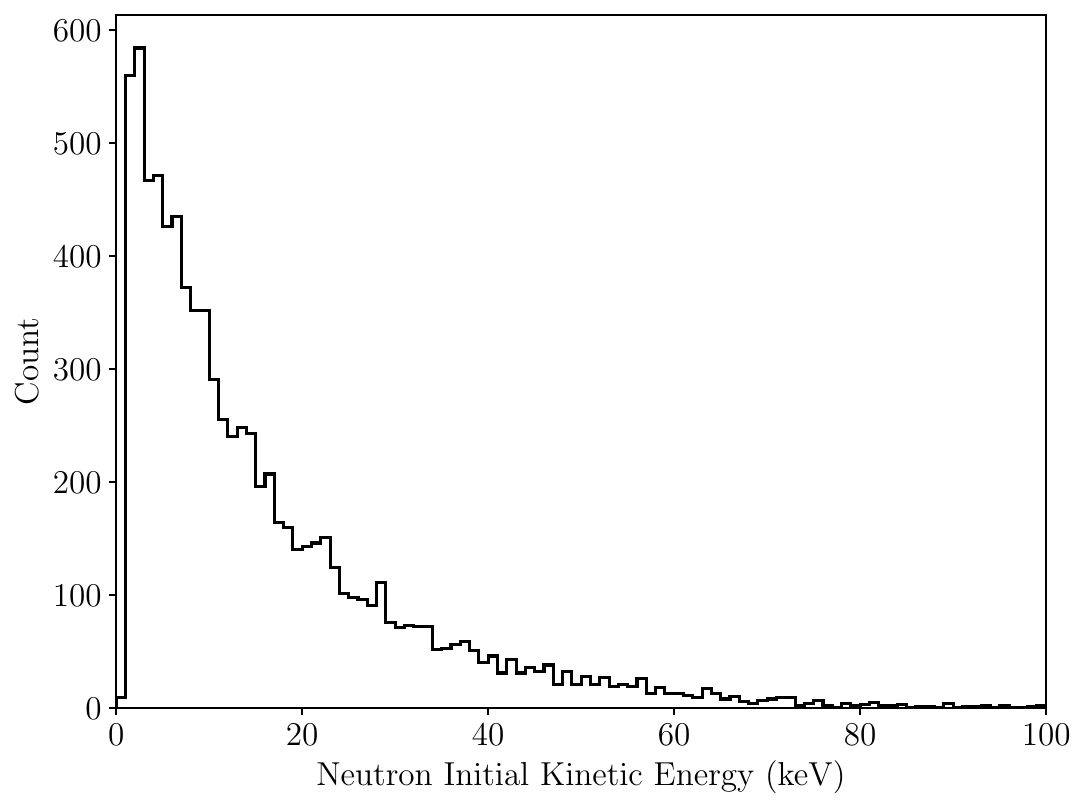}
    \caption{The initial kinetic energy spectrum of simulated IBD neutrons, filtered for $8,626$ captures on 0.1\%\ wt ${}^6\text{Li}$. The mean energy is $16\text{~keV}$, and standard deviation is $15\text{~keV}$.}
    \label{fig:neutron_spectrum}
\end{figure}

\begin{figure*}
    \centering
    \begin{overpic}[width=0.46\linewidth]
        {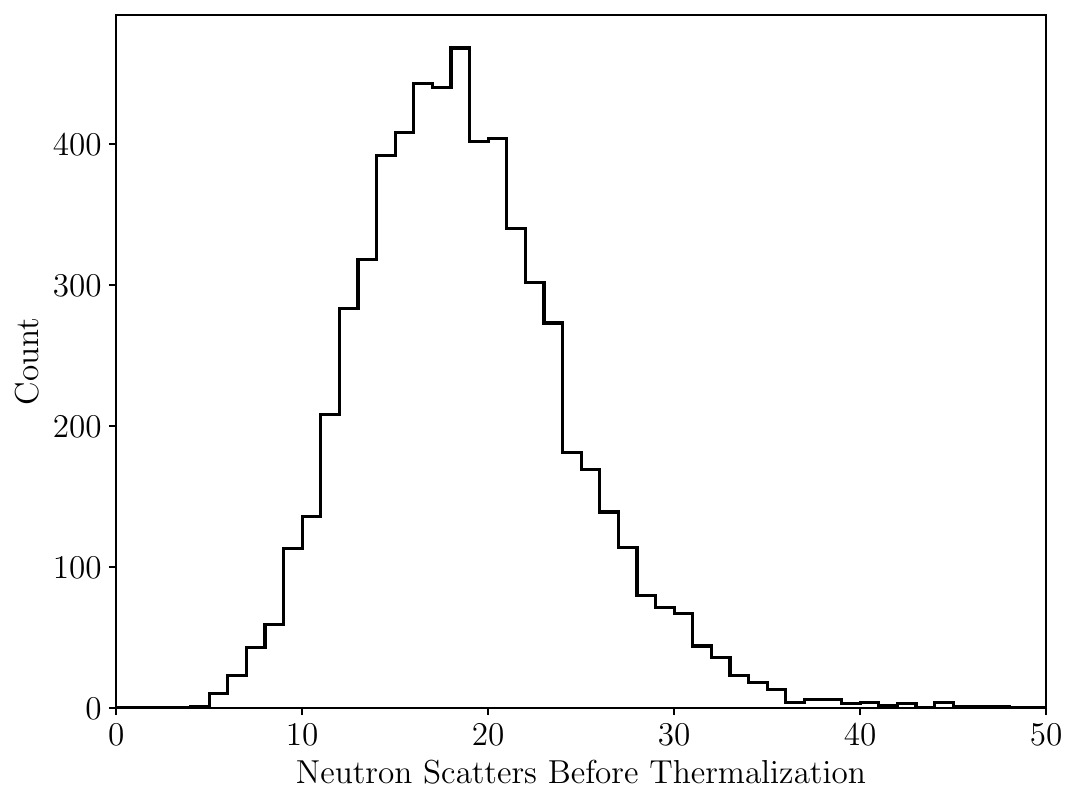}
    \end{overpic}
    \begin{overpic}[width=0.46\linewidth]
        {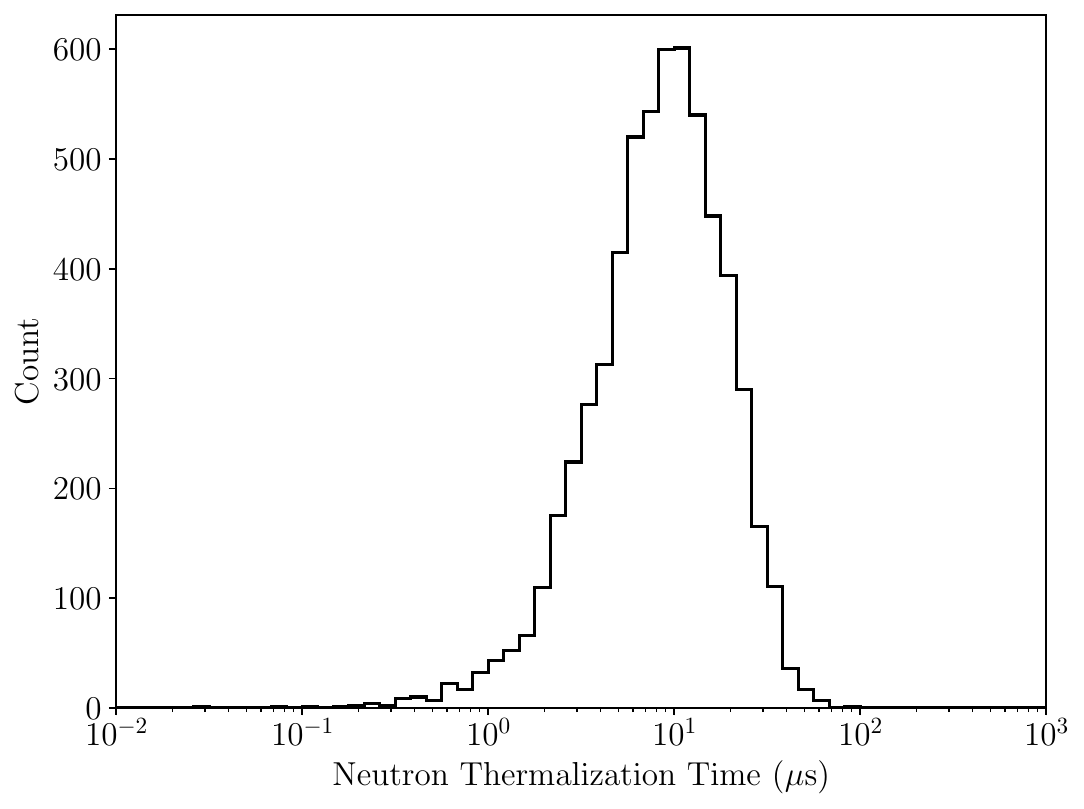}
    \end{overpic}\\
    \begin{overpic}[width=0.46\linewidth]
        {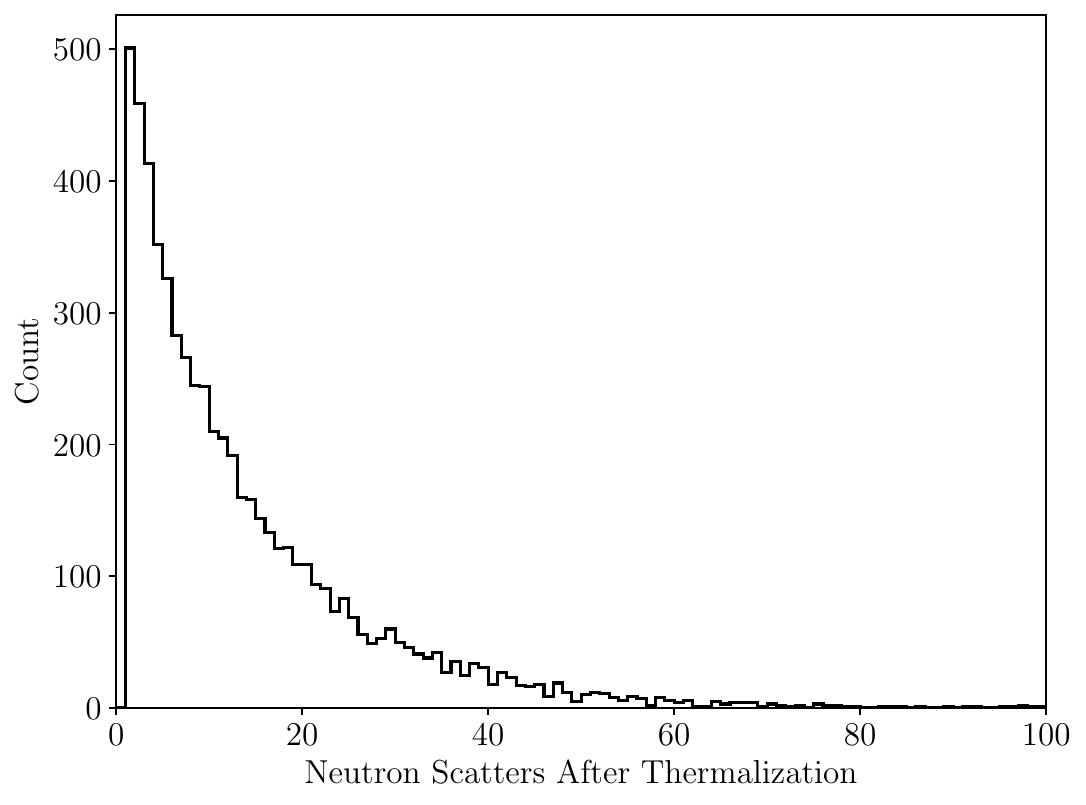}
    \end{overpic}
    \begin{overpic}[width=0.46\linewidth]
        {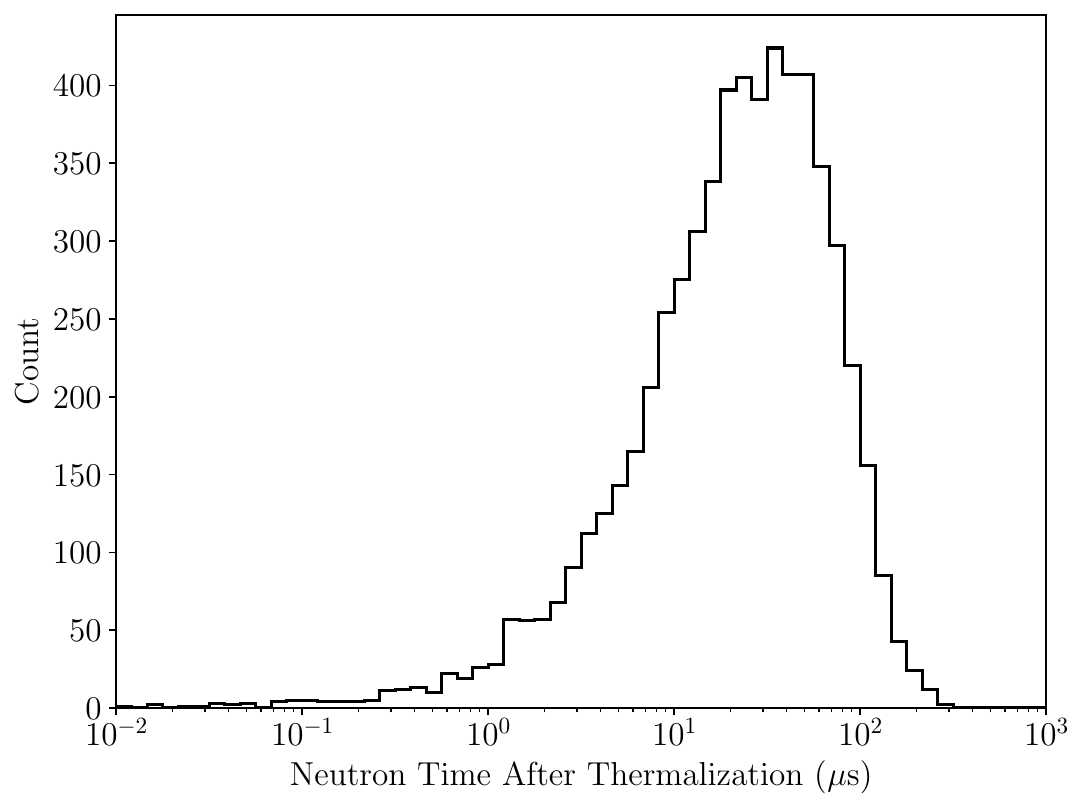}
    \end{overpic}
    \caption{Comparison of number of scatters (left) and time of flight (right) before (above) and after (below) thermalization, filtered for capture on 0.1\% by weight ${}^6\text{Li}$. Out of $10,000$ IBD neutrons, $8,626$ were captured on ${}^6\text{Li}$. Before thermalizing, neutrons scattered an average of 18 times with a standard deviation of 6. The average time to thermalize was 10 $\mu$s, with a standard deviation of 8 $\mu$s. After thermalization, neutrons scattered an average of 13 times with a standard deviation of 13. The average time between thermalization and capture was 34 $\mu$s, with a standard deviation of 34 $\mu$s.}
    \label{fig:neutronthermbeforeafter}
\end{figure*}

\begin{figure*}[ht]
    \centering
    \includegraphics[width=.49\linewidth]{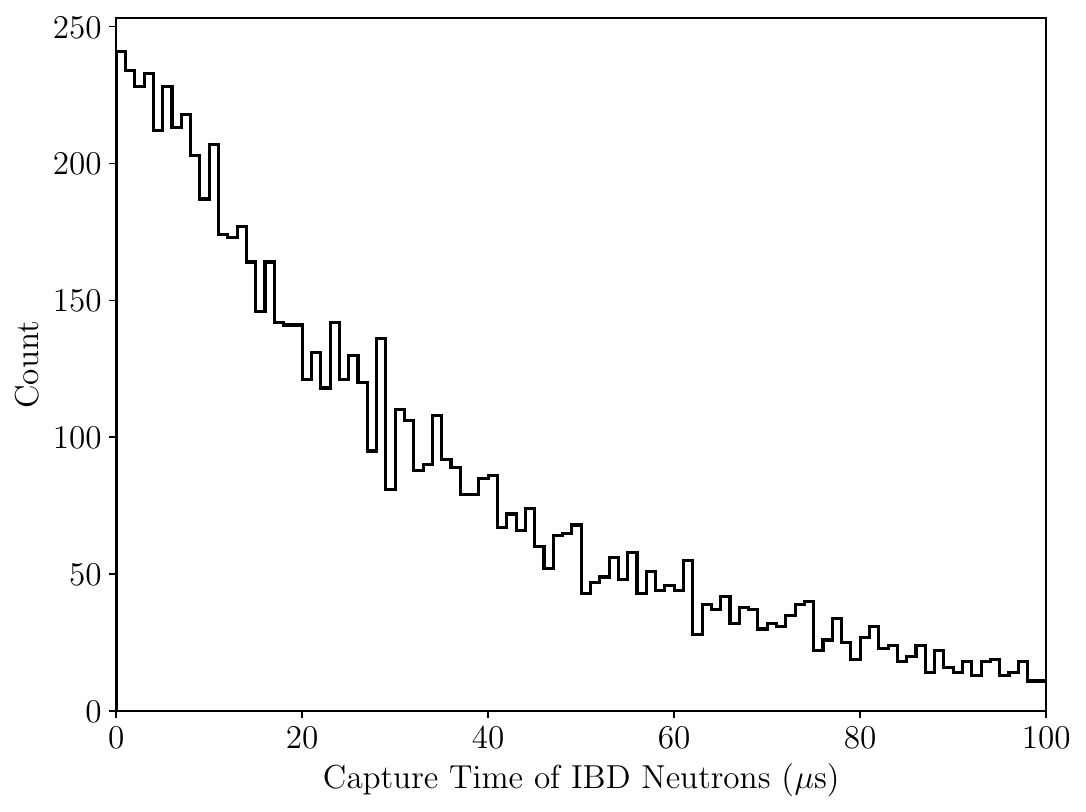}
    \includegraphics[width=.49\linewidth]{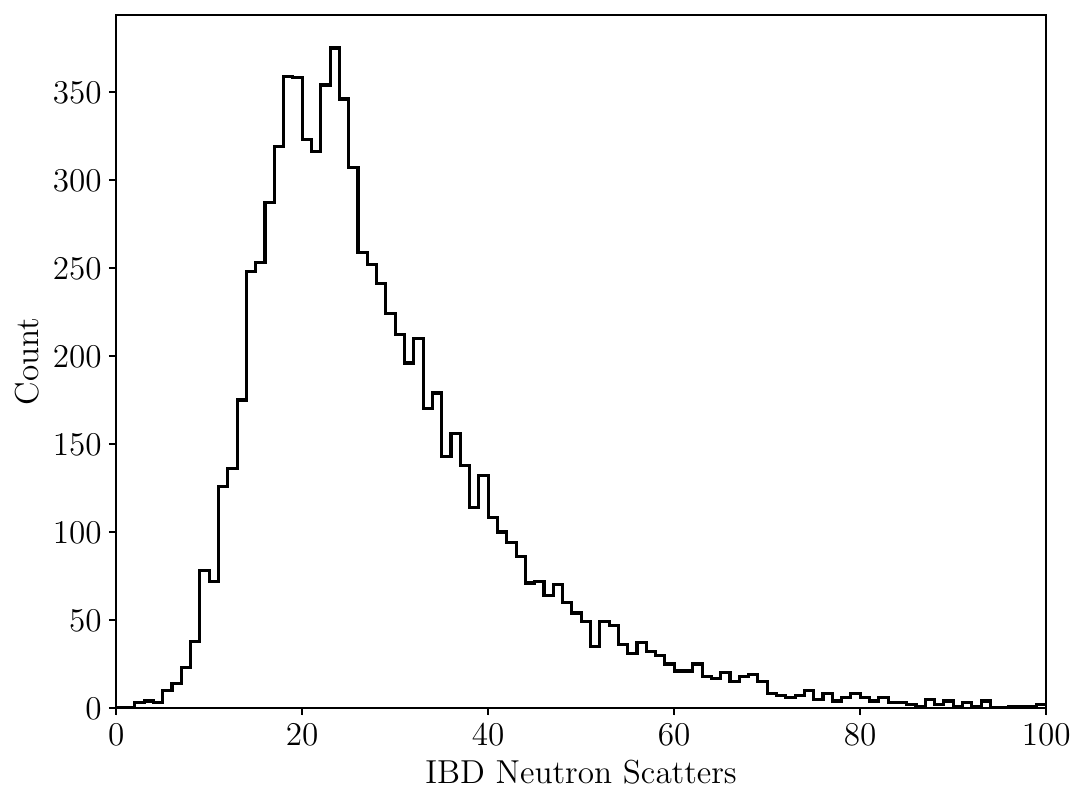}
    \caption{IBD-neutron capture time (left) and number of IBD-neutron scatters before capture (right), filtered for $8,626$ captures on ${}^6\text{Li}$, for 0.1\% by weight ${}^6\text{Li}$. The average capture time was 34~$\mu$s , with a standard deviation of 34~$\mu$s. The average number of scatters before capture was 28, with a standard deviation of 14.}
    \label{fig:neutroncapturetime_and_neutronscatters}
\end{figure*}

\begin{figure*}[ht]
    \centering
    \includegraphics[width=0.49\linewidth]{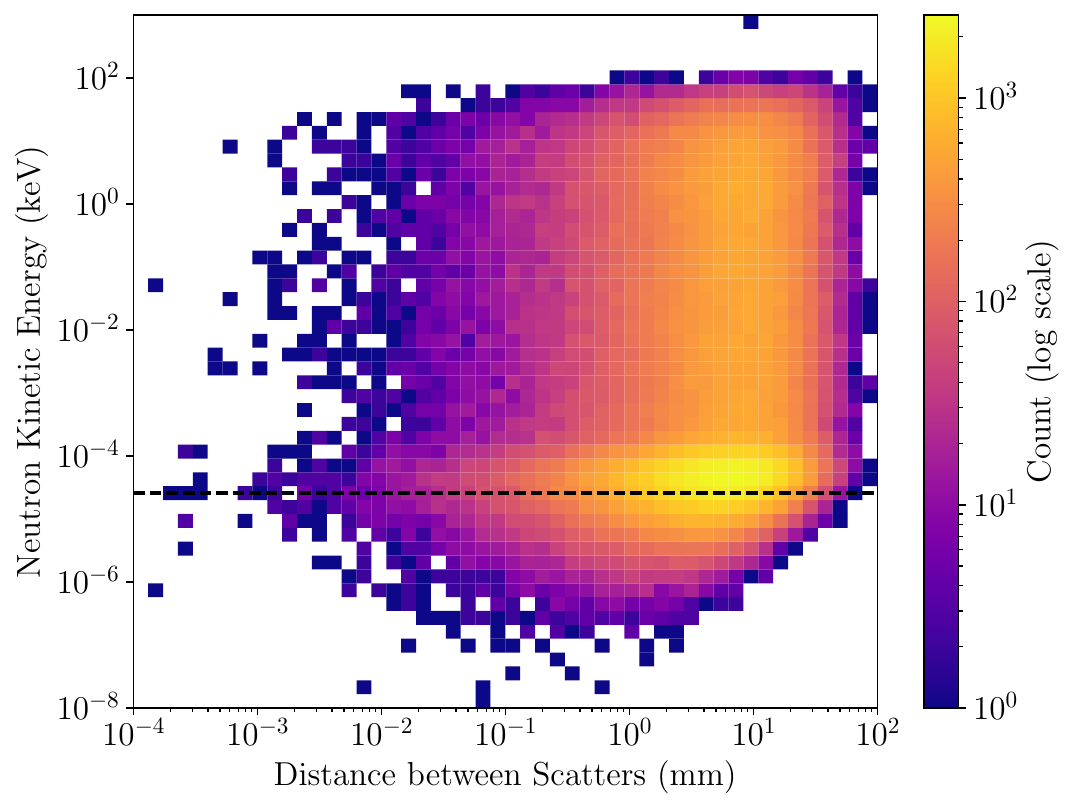}
    \includegraphics[width=0.49\linewidth]{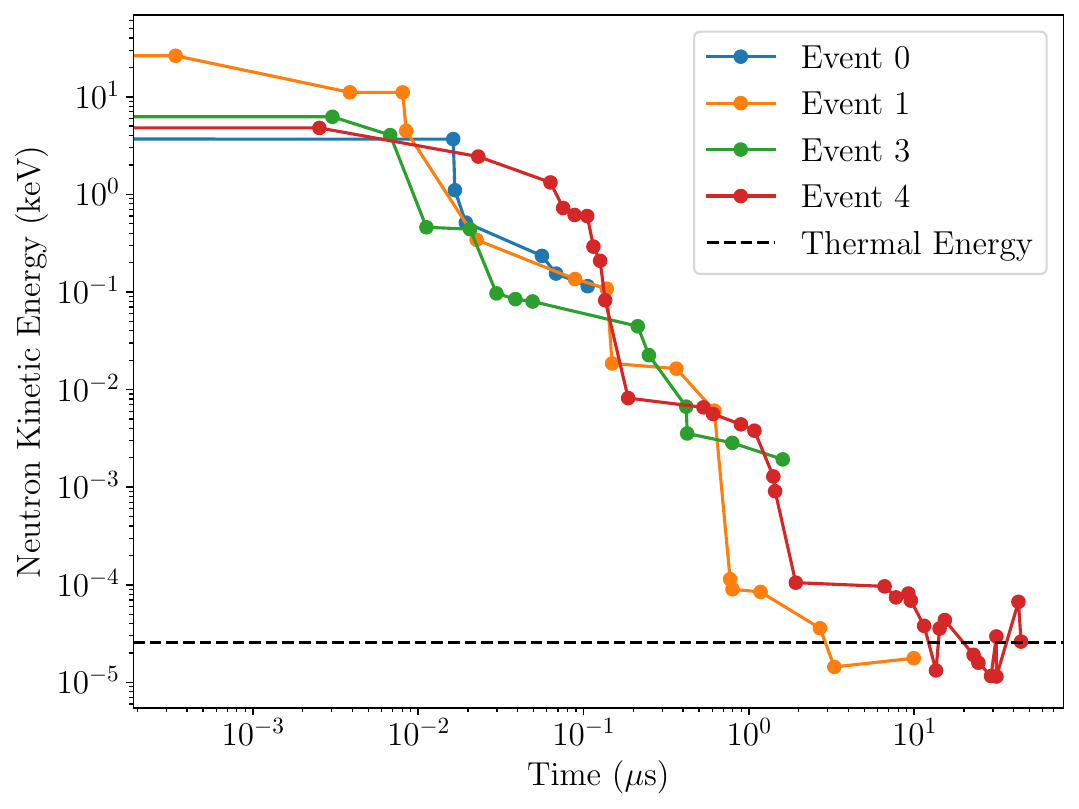}
    \caption{{\it Left:} Energy vs. distance between scatters in 0.1\%\ wt ${}^6$Li-doped scintillator. {\it Right:} Neutron energy vs time for selected IBD neutrons in 0.1\%\ ${}^6$Li doped scintillator. Events are filtered to include only captures on ${}^6$Li.  The dashed horizontal on both plots line represents thermal energy at 300 K.}
    \label{fig_neutron_scatters_energy_2D_hist_and_neutronenergyvtime}
\end{figure*}

A 10,000 IBD event dataset was used to validate the physics of the RAT-PAC 2 simulation. We made various histograms for IBD positrons and neutrons, reproduced here for clarity --- showing the positron and neutron track length for different dopant levels (Fig.~\ref{fig:ratpachistograms}), rendering of tracks (Figs.~\ref{fig:trackplot3dratpac}--\ref{fig:trackplot3dratpactopview}), the neutron kinetic energy spectrum (Fig. \ref{fig:neutron_spectrum}), neutron scatters before and after thermalization (Fig. \ref{fig:neutronthermbeforeafter}), neutron capture time and number of neutron scatters before capture (Fig. \ref{fig:neutroncapturetime_and_neutronscatters}), and a 2D histogram of neutron kinetic energy versus distance between scatters for the two doping concentrations we simulated, as well showing a neutron kinetic energy as a function of time for a few events (Fig.~\ref{fig_neutron_scatters_energy_2D_hist_and_neutronenergyvtime}). 

We decided that what qualified for a large dataset, for this study, was ten million total generated events for neutrinos (IBD positron-neutron pairs). Our reasoning is that from Fig.~\ref{fig:CFND_1iteration} and associated plots (presented in \cite{10.1063/5.0315079}), we had sufficient statistics at one million events, and that a factor of ten higher statistics was both statistically safe, and within reason for the computational cost to generate them.

\subsubsection{Parameter space}
Within simulating IBD events, there are many variables to consider. We construct a simple parameter space with the intention of further expanding upon it as we work to incorporate realistic backgrounds and detector materials and components. For the scope of this publication, which  prioritizes presenting an improved directionality and uncertainty algorithm, we can describe these parameters $p$ as a vector $\vec{p} = \{p_i|i\in\mathbb{N}\}$, where $p_1,p_2 = \theta_1,\theta_2$ $p_3 = L$, and $p_{4+}$ = $^6$Li concentration, segment size.

For the beginning of this study we are starting with a 2-d Euclidean space in X-Y, and determining the relative direction $\theta$. We are also only simulating a singular source that produces a planar-wave neutrino wind, meaning there is no purpose in keeping L for interaction or oscillating properties. Since our detector is "fixed", it means that we will not vary segment size, dopant level, number of events, or any other parameters related to constant constraints as that will be held constant between simulation and detector and thus will not change. This leaves only $\theta$. As complexity in our simulations increases, we will need to include more parameter, however at this stage in simulation, the other parameters do not change anything.

\newpage
\bibliography{ref.bib}
\bibliographystyle{aipnum4-2}

\end{document}